\documentclass[conference]{IEEEtran}
\IEEEoverridecommandlockouts
\usepackage{cite}
\usepackage{amsmath,amssymb,amsfonts}
\usepackage{algorithmic}
\usepackage{graphicx}
\usepackage{textcomp}
\usepackage{xcolor}
\usepackage{url}
\usepackage{bm}
\usepackage{multirow}
\usepackage{makecell}
\usepackage[square,comma,numbers,sort]{natbib}
\usepackage[nomargin,inline,marginclue,draft]{fixme}
\usepackage{threeparttable}
\usepackage{booktabs}
\usepackage{subfig}
\usepackage[numbers]{natbib}

\def\BibTeX{{\rm B\kern-.05em{\sc i\kern-.025em b}\kern-.08em
    T\kern-.1667em\lower.7ex\hbox{E}\kern-.125emX}}

\begin{document}

\title{Cross-Domain Recommendation to Cold-Start Users via Variational Information Bottleneck
}

\author{
    \IEEEauthorblockN{Jiangxia Cao\IEEEauthorrefmark{2}\IEEEauthorrefmark{3},
    Jiawei Sheng\IEEEauthorrefmark{2}\IEEEauthorrefmark{3},
    Xin Cong\IEEEauthorrefmark{2}\IEEEauthorrefmark{3},
    Tingwen Liu\IEEEauthorrefmark{2}\IEEEauthorrefmark{3}\IEEEauthorrefmark{1}, 
    Bin Wang\IEEEauthorrefmark{4}}
    \IEEEauthorblockA{\IEEEauthorrefmark{2} Institute of Information Engineering, Chinese Academy of Sciences, Beijing, China}
    \IEEEauthorblockA{\IEEEauthorrefmark{3} School of Cyber Security, University of Chinese Academy of Sciences, Beijing, China}
    \IEEEauthorblockA{\IEEEauthorrefmark{4} Xiaomi AI Lab, Beijing, China}
    \thanks{\IEEEauthorrefmark{1}Corresponding Author}
Email: 
\IEEEauthorrefmark{2}\IEEEauthorrefmark{3}\{caojiangxia, shengjiawei, congxin, liutingwen\}@iie.ac.cn,
\IEEEauthorrefmark{4}wangbin11@xiaomi.com
}
\maketitle

\begin{abstract}
Recommender systems have been widely deployed in many real-world applications, but usually suffer from the long-standing user cold-start problem. 
As a promising way, Cross-Domain Recommendation (CDR) has attracted a surge of interest, which aims to transfer the user preferences observed in the source domain to make recommendations in the target domain.
Previous CDR approaches mostly achieve the goal by following the Embedding and Mapping (EMCDR) idea which attempts to learn a mapping function to transfer the pre-trained user representations (embeddings) from the source domain into the target domain.
However, they pre-train the user/item representations independently for each domain, ignoring to consider both domain interactions simultaneously.
Therefore, the biased pre-trained representations inevitably involve the domain-specific information which may lead to negative impact to transfer information across domains. 
In this work, we consider a key point of the CDR task: what information needs to be shared across domains?
To achieve the above idea, this paper utilizes the information bottleneck (IB) principle, and proposes a novel approach termed as CDRIB to enforce the representations encoding the domain-shared information.
To derive the unbiased representations, we devise two IB regularizers to model the cross-domain/in-domain user-item interactions simultaneously and thereby CDRIB could consider both domain interactions jointly for de-biasing.
With an additional contrastive information regularizer, CDRIB can also capture cross-domain user-user correlations.
In this way, those regularizers encourage the representations to encode the domain-shared information, which has the capability to make recommendations in both domains directly.
To the best of our knowledge, this paper is the first work to capture the domain-shared information for cold-start users via variational information bottleneck.
Empirical experiments illustrate that CDRIB outperforms the state-of-the-art approaches on four real-world cross-domain datasets, demonstrating the effectiveness of adopting the information bottleneck for CDR. 
\end{abstract}

\begin{IEEEkeywords}
Cross-Domain Recommendation; User Cold-Start Recommendation; Information Bottleneck
\end{IEEEkeywords}

\section{Introduction}

To alleviate information overload on the web, Recommender Systems (RS) have been widely deployed for personalized information filtering in many real-world applications such as Amazon (E-commerce) and Youtube (online video).
As the cornerstone of RS, collaborative filtering (CF) achieves great success in understanding user preferences~\cite{neumf,bvcf}, which learns user/item representations based on the observed user-item interactions.
However, these CF-based methods usually suffer from the long-standing user cold-start problem~\cite{lincsr2021}, i.e., new coming users have non-observed interactions to learn effective representations for the recommendation.

\begin{figure}[t]
	\begin{center}
		\includegraphics[width=8.5cm]{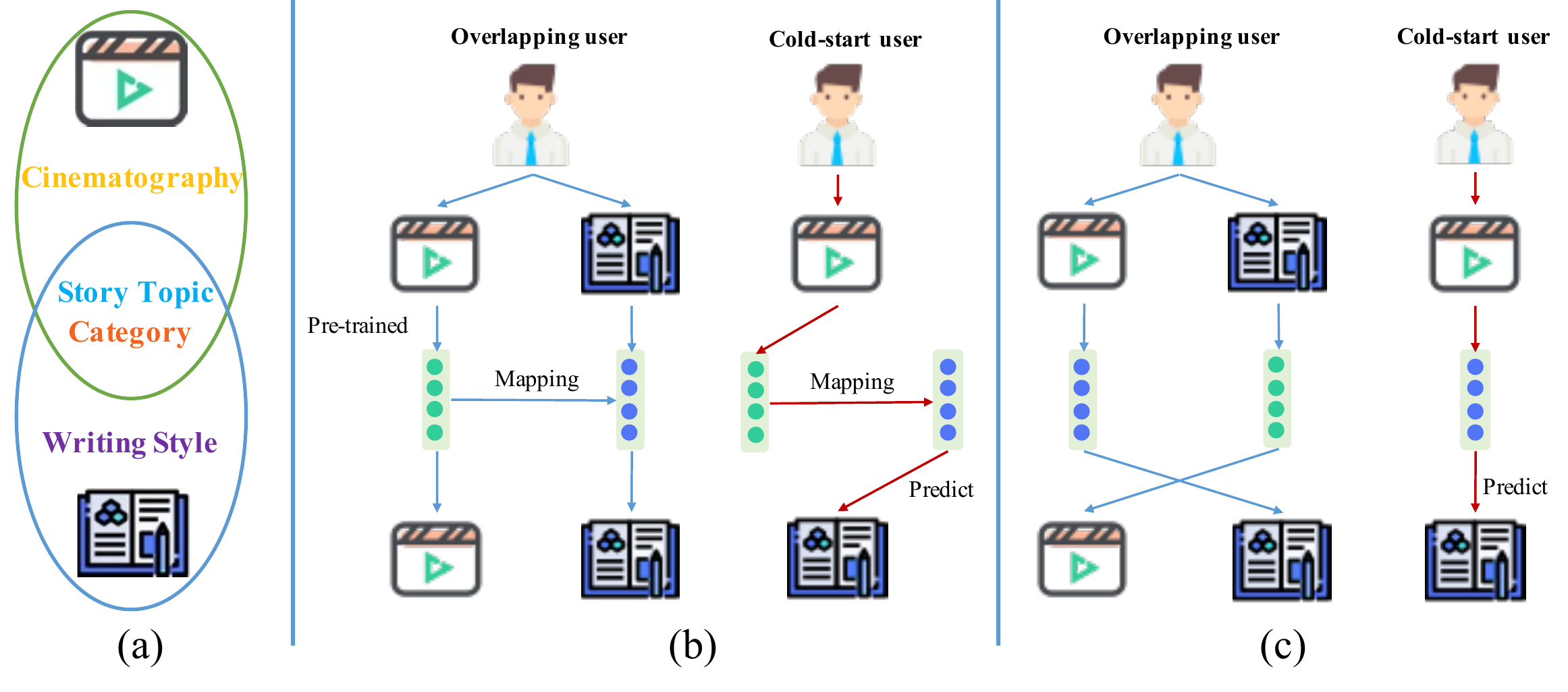}
		\caption{(a): An example of preferences in \texttt{Film} and \texttt{Book} domain.
			(b): The pipeline framework in EMCDR-based methods by pre-training/mapping embeddings.
			(c): The joint framework in our CDRIB by jointly learning cross-domain embeddings. 
			The overlapping users bridge both domains in CDR, while our task is to make recommendations for the cold-start users.
			} 
		\label{motivation}
	\end{center}
\end{figure}
To mitigate the user cold-start problem, cross-domain recommendation (CDR)~\cite{crosssurvey} has attracted much attention in recent years.
Generally, CDR aims to improve recommendation quality for the target domain by leveraging the user-item interaction information collected from other relative source domains.
In this task, CDR requires a number of users existing in both domains (i.e., overlapping users) to bridge the source and target domain.
Those users only existed in the source domain (i.e., non-overlapping users) can be seen as the \textbf{cold-start users} in the target domain.
To deal with the challenging user cold-start problem, recent CDR approaches mostly follow the Embedding and Mapping idea (EMCDR~\cite{emcdr}).
Fig.~\ref{motivation}(b) gives the illustration of the EMCDR-based methods.
Specifically, EMCDR first generates user/item representations with CF-based models for each domain separately, and then trains a mapping function to align the overlapping user representations between the source domain and the target domain.
Afterward, given a non-overlapping (cold-start) user representation from the source domain, it can predict the user representation in the target domain by the trained mapping function, and then recommend items in the target domain.

Despite their success, most EMCDR-based approaches~\cite{emcdr,sscdr,tmcdr,recsys,vcross} pre-train user/item representations independently which could be easily biased on each domain, limiting the transferring effectiveness of the mapping function.
As shown in Fig.~\ref{motivation}(a), there exist several kinds of user preference in two domains: \texttt{Film} and \texttt{Book}.
On the one hand, the information shared across domains (such as \texttt{Story Topic} and \texttt{Category} in both domains) usually provides valuable clues for CDR.
For example, a user preferring film with the romantic category may also prefer the same category when buying books.
On the other hand, other domain-specific information (such as \texttt{Cinematography} in \texttt{Film} and \texttt{Writing Style} in \texttt{Book} domain) may provide useless clues, even causing the negative transfer problem~\cite{conet}.
For instance, the preference for 3D film may be useless to recommend books for the users.
Unfortunately, the pre-trained representations inevitably encode the domain-specific information.

We argue that it can be beneficial to jointly learn user/item representations by considering both domain interactions, further focusing on the domain-shared information and limiting the domain-specific information.
To this end, we introduce the employed technique, information bottleneck~(IB)~\cite{orignalib,ib} from information theory, to generate effective user/item representations for CDR.
Generally, IB principle aims at deriving ``good" representations, in terms of \textit{a fundamental tradeoff between having a concise representation with its own information and one with general predictive power}~\cite{vib}.
In our task, we encourage the representations debiased from their source domain, and remain predictive in the target domain, which also introduces a tradeoff according to the IB principle.
With the variational inference to solve the intractable mutual information in IB, we could derive effective representations which focus on encoding the domain-shared information to make recommendations in both domains directly.

\begin{figure*}[t]
	\begin{center}
		\includegraphics[width=16cm,height=7cm]{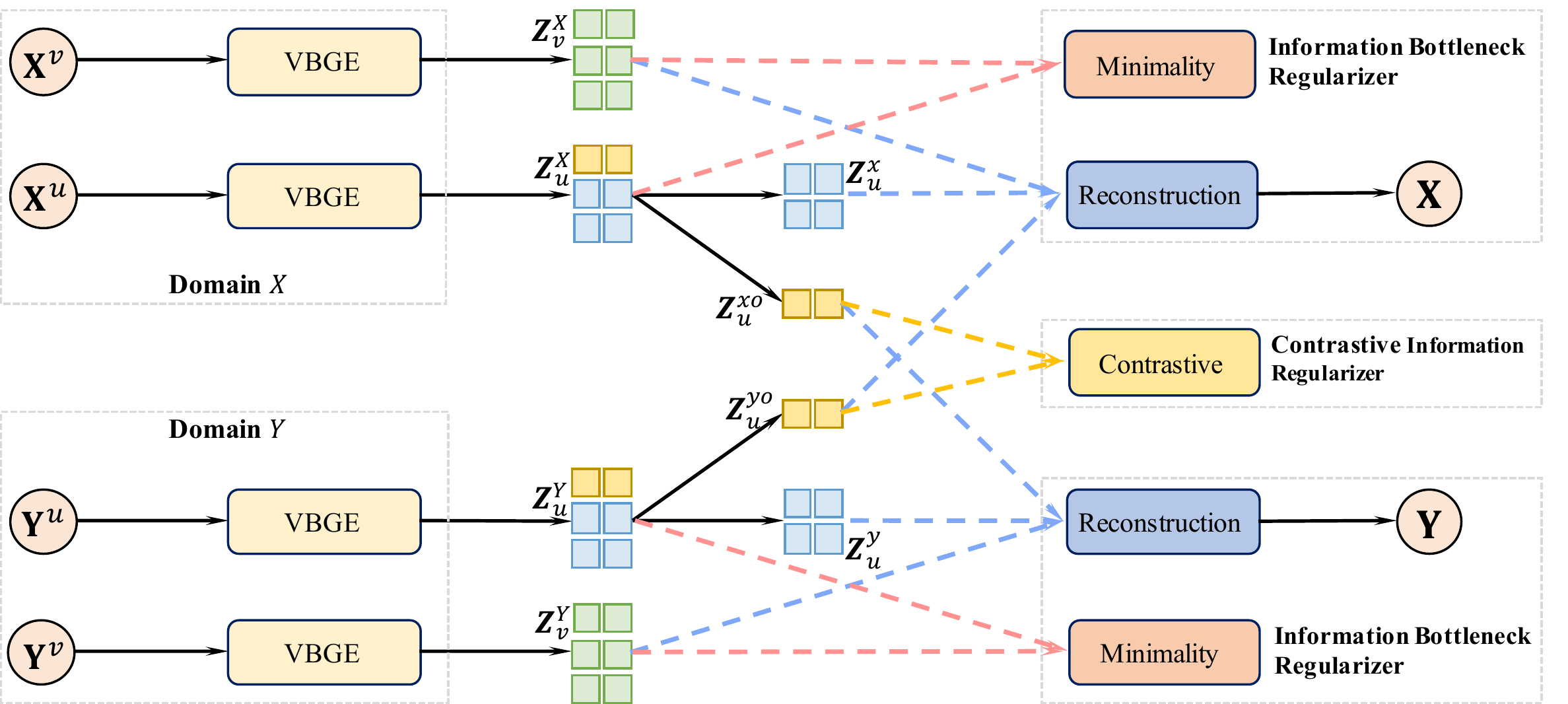}
		\caption{The overview of CDRIB, which contains VBGEs to generate user/item representations, and two kinds of regularizers to build cross-domain correlations. The green colored grids denote the item representations. The yellow and blue colored grids denote the overlapping and non-overlapping user representations. The used notations are detailed in TABLE~\ref{tab::symbol}.
		}
		\label{totalmodel}
	\end{center}
\end{figure*}

To implement the above idea, we propose a novel variational approach for \underline{C}ross-\underline{D}omain \underline{R}ecommendation to cold-start users via variational \underline{I}nformation \underline{B}ottleneck termed as \textbf{CDRIB} (As shown in Fig.~\ref{motivation}(c) and Fig.~\ref{totalmodel}).
Specifically, we first devise a variational bipartite graph encoder (VBGE) to generate user/item representations in each domain as a fundamental encoder.
To model the user-item interactions in both domains jointly and debias the user/item representations from their source domain, we devise two \textbf{information bottleneck regularizers} for the cross-domain and in-domain user-item interactions respectively.
The former information bottleneck regularizer learns user/item representations with the cross-domain user-item interactions (i.e., the interactions of overlapping users observed in domain $X$ and $Y$), while the latter information bottleneck regularizer learns user/item representations with the in-domain user-item interactions.
By optimizing the IB regularizers, CDRIB encourages user/item representations to encode domain-shared information and limit domain-specific information, and naturally has the capability to make cross-domain recommendations to cold-start users.
Additionally, to exploit the user-user correlations across domains, we design a \textbf{contrastive information regularizer} over the overlapping users and thus the user representations are further aligned to enhance the domain-shared information.

In summary, the contributions of this paper are as follows:
\begin{itemize}
	\item This paper introduces a fresh perspective to solve CDR by learning user/item representations of two domains jointly to capture the domain-shared information.

	\item This paper devises a novel approach, CDRIB, which contains two information bottleneck regularizers to build user-item interactions across domains and a contrastive information regularizer to align user representations overlapped in different domains.

    \item Empirical experiments demonstrate that our CDRIB outperforms the state-of-the-art approaches on 4 real-world cross-domain datasets with a raising of 34.38\% averagely in terms of MRR. Besides, we also conduct extensive ablation studies and detailed analyses to verify the effectiveness of our method.
    Our codes and datasets are available for further comparison\footnote{\url{https://github.com/cjx96/CDRIB}}.
\end{itemize}

\begin{table}[t]
	\centering
	\caption{A list of commonly used notations.}
	\setlength\tabcolsep{4pt}{
	\begin{tabular}{cl}
		\toprule
		\textbf{Notation} & \textbf{Description}\\
		\midrule
		$\mathcal{U}^X$, $\mathcal{U}^Y$ & The user set. \\ \specialrule{0em}{2pt}{2pt}
		$\mathcal{V}^X$, $\mathcal{V}^Y$ & The item set. \\ \specialrule{0em}{2pt}{2pt}
		$\mathcal{U}^x$, $\mathcal{U}^y$ & The non-overlapping (cold-start) user set. \\ \specialrule{0em}{2pt}{2pt}
		$\mathcal{U}^o$ & The overlapping user set across domains. \\ \specialrule{0em}{2pt}{2pt}
		$\bm{A}^X$, $\bm{A}^Y$ & The binary adjacent matrix for user-item interaction. \\ \specialrule{0em}{2pt}{2pt}
		\midrule
		$\bm{Z}_u^X$, $\bm{Z}_u^Y$ & User representations. \\ \specialrule{0em}{2pt}{2pt}
        $\bm{Z}_v^X$, $\bm{Z}_v^Y$ & Item representations. \\ \specialrule{0em}{2pt}{2pt}
		$\bm{Z}_u^{xo}$, $\bm{Z}_u^{yo}$ & Overlapping user representations.\\ \specialrule{0em}{2pt}{2pt}
        $\bm{Z}_u^{x}$, $\bm{Z}_u^{y}$ & Non-overlapping (cold-start) user representations.\\ \specialrule{0em}{2pt}{2pt}
		$\mathbf{X}$, $\mathbf{Y}$ & The user-item interaction information. \\ \specialrule{0em}{2pt}{2pt}
        $\mathbf{X}^u$, $\mathbf{Y}^u$ & The user information (users' features and structure). \\ \specialrule{0em}{2pt}{2pt}
		$\mathbf{X}^v$, $\mathbf{Y}^v$ & The item information (items' features and structure). \\ \specialrule{0em}{2pt}{2pt}
		\midrule
		$\mathcal{L}_{o2X}$, $\mathcal{L}_{o2Y}$ & Cross-domain information bottleneck regularizer. \\ \specialrule{0em}{2pt}{2pt}
		$\mathcal{L}_{x2X}$, $\mathcal{L}_{y2Y}$ & In-domain information bottleneck regularizer. \\ \specialrule{0em}{2pt}{2pt}
		$\mathcal{L}_{con}$ & Contrastive information regularizer. \\ \specialrule{0em}{2pt}{2pt}
	\bottomrule
	\end{tabular}
	\label{tab::symbol}
	}
\end{table}

\section{Preliminaries}
\subsection{Problem Formulation}

This work considers a general CDR scenario where there exist two domains $X$, $Y$ with users and items.
Let $D^X = (\mathcal{U}^X, \mathcal{V}^X, \mathcal{E}^X)$ and $D^Y = (\mathcal{U}^Y, \mathcal{V}^Y, \mathcal{E}^Y)$ denote the domain data, where $\mathcal{U}$, $\mathcal{V}$ and $\mathcal{E}$ are user set, item set and edge set in each domain respectively.
Particularly, the user set $\mathcal{U}^X$ and $\mathcal{U}^Y$ contain an overlapping user subset $\mathcal{U}^o = \mathcal{U}^X \bigcap \ \mathcal{U}^Y$. Then, the user sets can be formulated as $\mathcal{U}^X = \{\mathcal{U}^x, \mathcal{U}^o\}$ and $\mathcal{U}^Y = \{\mathcal{U}^y, \mathcal{U}^o\}$, where $\mathcal{U}^x$ and $\mathcal{U}^y$  are the non-overlapping user set in each domain.
For simplicity, we further introduce two binary matrices to store the user-item interactions, i.e., $\bm{A}^X = \{0,1\}^{|\mathcal{U}^X|\times|\mathcal{V}^X|}$ and $\bm{A}^Y = \{0,1\}^{|\mathcal{U}^Y|\times|\mathcal{V}^Y|}$, where the element $A_{ij}$ in each domain denotes whether the user $u_i \in \mathcal{U}$ and item $v_j \in \mathcal{V}$ have an interaction in the edge set $\mathcal{E}$.


Given the observed interaction data of both domains, CDR aims to make recommendations in the target domain for non-overlapping (cold-start) users who are only observed in the source domain.
Formally, given a non-overlapping (cold-start) user $u_i \in \mathcal{U}^x$ from source domain $X$, we would like to recommend item $v_j \in \mathcal{V}^Y$ from target domain $Y$ (vice versa in the case of users from $\mathcal{U}^y$ and items from $\mathcal{V}^X$).

\subsection{Information Bottleneck Principle}
The information bottleneck was first proposed by~\citeauthor{orignalib}~\cite{orignalib,ib}, which aims at learning effective representations with a tradeoff between having a concise representation and one with general predictive power~\cite{vib}.
Formally, the standard information bottleneck has the following objective function:
\begin{equation}
	\mathcal{L}_{IB} := \beta I(\bm{Z};\mathbf{X}) - I(\bm{Z};\mathbf{Y}).
\end{equation}
Here the $I(\cdot;\cdot)$ denotes the mutual information of two random variables.
The goal is to learn an optimal representation $\bm{Z}$ that is maximally expressive about $\mathbf{Y}$ while minimally relies on $\mathbf{X}$, where the $\beta> 0$ is a Lagrangian multiplier to control the trade-off.
Intuitively, minimizing the above objective function can be interpreted as: (1) Minimizing the first term $I(\bm{Z};\mathbf{X})$ aims to penalize the information between $\bm{Z}$ and $\mathbf{X}$. That is to say, the latent variable $\bm{Z}$ will somewhat ``forget'' information of $\mathbf{X}$. 
(2) Maximizing the last term $I(\bm{Z};\mathbf{Y})$ encourages $\bm{Z}$ to be predictive of $\mathbf{Y}$.

Actually, the IB principle forces the representation $\bm{Z}$ to capture the relevant factors and compress $\mathbf{X}$ by diminishing the irrelevant parts which do not contribute to the prediction of $\mathbf{Y}$~\cite{mine}.
That is to say, IB encourages $\bm{Z}$ to act as a \textit{minimal sufficient statistic}~\cite{gib} (i.e., domain-shared information in our CDR task).
In practice, directly optimizing the mutual information is intractable, thereby the variational approximation~\cite{amortized} is widely used to construct a lower bound to optimize the information bottleneck objective function~\cite{vib,cvib}.

\section{Methodology}

This section introduces the components of our model CDRIB.
Fig.~\ref{totalmodel} sketches the overview of CDRIB, including a variational bipartite graph encoder (VBGE) and two kinds of cross-domain information regularizers.
We start from the embedding layer, which provides initialized user/item representations for the following components.
Then, VBGE builds the user-item interactions as a bipartite graph, and generates user/item latent variables (a.k.a. representations) with variational inference for each domain.
Afterward, two kinds of regularizers are proposed to constrain cross-domain correlation to enforce the representation encoding the domain-shared information.
The information bottleneck regularizer builds the cross-domain and in-domain user-item interactions, while the contrastive information regularizer captures cross-domain user-user correlations.
Finally, we give details of tractable objective functions to optimize the overall model.

\subsection{Embedding Layer}

The embedding layer provides initialized representations for users and items in each domain.
Specifically, it contains four parameter matrices as embeddings, including $\bm{U}^X \in \mathbb{R}^{|\mathcal{U}^X| \times F}$ and $\bm{V}^X\in \mathbb{R}^{|\mathcal{V}^X| \times F}$ for user/item in domain $X$, and $\bm{U}^Y\in \mathbb{R}^{|\mathcal{U}^Y| \times F}$ and $\bm{V}^Y\in \mathbb{R}^{|\mathcal{V}^Y| \times F}$ for those in domain $Y$ respectively, where $F$ is the representation dimension.

\subsection{Variational Bipartite Graph Encoder}

To generate user/item representations, this section proposes the variational bipartite graph encoder (VBGE).
Different from general homogeneous graphs~\cite{homohete}, the user-item interactions can be regarded as a heterogeneous bipartite graph~\cite{gcmc,cao2021deep}, where any two user nodes are connected with the even-number-hop (e.g., 2-hop, 4-hop, etc).
Nevertheless, traditional graph encoders~\cite{gcn,ngcf,pinsage} mostly aggregate node information from its direct (1-hop) neighbors, which does not explicitly aggregate user/item node from its homogeneous neighbors.
Inspired by BiGI~\cite{bigi}, our VBGE exploits a two-step learning procedure, which first obtains interim representation with homogeneous even-number-hop neighbors, and then generates latent variables (a.k.a, representations) with the variational inference framework~\cite{vae}.
In this way, user/item representations could aggregate information from its homogeneous neighbors.

\begin{figure}[t]
 	\begin{center}
 		\includegraphics[width=8.5cm,height=4cm]{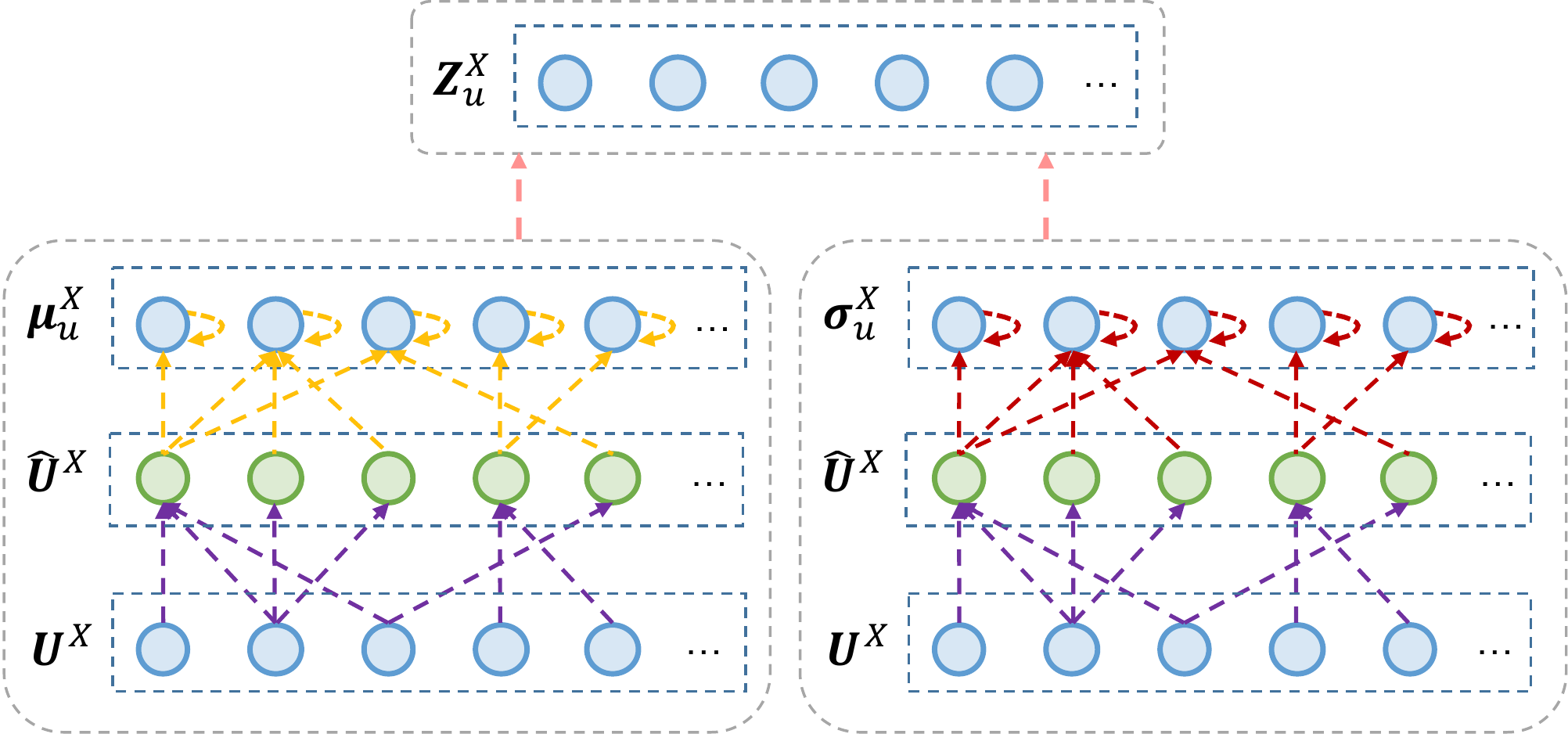}
 		\caption{The learning procedure of our VBGE to generate user representations $\bm{Z}_{u}^{X}$. $\widehat{\bm{U}}^X$ denotes the interim representations. The blue and green circles denote user and item nodes. The purple dotted lines denote Eq.(\ref{vbge_1}). The yellow, red and pink dotted lines denote Eq.(\ref{vbge_2}).}
 		\label{vbge}
 	\end{center}
 \end{figure}

For simplicity, the following section takes the learning procedure of user representations as an example (Fig.~\ref{vbge} gives an illustration).
In order to connect users with their even-number-hop user neighbors, the first step of VBGE generates user interim representations to aggregate homogeneous information.
For instance, taking the users of domain $X$ as input, their interim representations $\widehat{\bm{U}}^X$ can be generated as follows:
\begin{equation}
	\small
	\begin{split}
		\widehat{\bm{U}}^X = \delta & \Big(\texttt{Norm}\big((\bm{A}^X)^\top\big) \bm{U}^X\bm{W}^X_u\Big), \\
	\end{split}
	\label{vbge_1}
\end{equation}
where $\texttt{Norm}(\cdot)$ means the row-normalized function, $\delta(\cdot)$ is the LeakyReLU function\footnote{Though recent studies~\cite{lightgcn, sgc} empirically demonstrate that the nonlinear factors in GNNs may have negative effect to the recommendation task, we still find that the activation function works in our experiments. The similar phenomenon also appears in Eq.~(\ref{vbge_2}).}, and $\bm{W}^X_u$ is a parameter matrix. Note that $(\bm{A}^X)^\top\in \mathbb{R}^{|\mathcal{V}^X|\times|\mathcal{U}^X|}$ is the transposed adjacent matrix, reflecting the directed connections from user nodes to item nodes.
In this way, the interim representation $\widehat{\bm{U}}^X \in \mathbb{R}^{|\mathcal{V}^X|\times F}$ is generated from user neighbors, only aggregating the homogeneous information of users.

Based on the interim representations, the second step of VBGE further aggregates interim representation to generate latent variables for the cross-domain information regularizer (in Section~\ref{Sec.Regularizer}). 
Specifically, the user latent variables $\bm{Z}^X_u$ can be obtained as follows:
\begin{equation}
	\small
	\begin{split}
		\bm{\mu}^X_u = \delta \Big( \Big[ \delta (\texttt{Norm}&(\bm{A}^X)\widehat{\bm{U}}^X\widehat{\bm{W}}^X_{u,\mu}) \oplus \bm{U}^X \Big] \bm{W}^X_{u,\mu} \Big),\\
		\bm{\sigma}^X_u = \varphi \Big(\big[\delta(\texttt{Norm}&(\bm{A}^X)\widehat{\bm{U}}^X\widehat{\bm{W}}^X_{u,\sigma}) \oplus \bm{U}^X\big]\bm{W}^X_{u,\sigma}\Big),\\
		\bm{Z}_u^X \sim& \mathcal{N}\Big(\bm{\mu}^X_u, [\text{diag}(\bm{\sigma}^X_u)]^2\Big),
	\end{split}
	\label{vbge_2}
\end{equation}
where $\oplus$ is the concatenation operation, $\varphi(\cdot)$ is the Softplus function, and $\widehat{\bm{W}}^X_{u,\mu}, \bm{W}^X_{u,\mu}, \widehat{\bm{W}}^X_{u,\sigma}, \bm{W}^X_{u,\sigma}$ are parameter matrices.
Note that $\bm{A}^X \in \mathbb{R}^{|\mathcal{U}^X|\times|\mathcal{V}^X|}$ reflects the directed connections from item nodes to user nodes.
It is also worth noting that $\bm{\mu}^X_u$ and $\bm{\sigma}^X_u$ are mean and standard deviation of multivariate Gaussian distribution $\mathcal{N}(\cdot,\cdot)$, which is used to sample the latent variables $\bm{Z}_u^X\in \mathbb{R}^{|\mathcal{U}^X|\times F}$.
For implementation, we adopt the widely used reparameterization trick~\cite{vae} to sample the user representation $\bm{z}_{u_i}^X$ as:
\begin{equation}
	\small
	\begin{split}
		\bm{z}_{u_i}^X = \bm{\mu}^X_{u_i} + &\bm{\sigma}^X_{u_i} \odot \bm{\epsilon}, \quad \bm{\epsilon} \sim \mathcal{N}\Big( 0,\text{diag}(\bm{I}) \Big), \\
	\end{split}
	\label{repara}
\end{equation}
where $\bm{\mu}^X_{u_i}$ and $\bm{\sigma}^X_{u_i}$ are mean and standard deviation vector of user $u_i$ in domain $X$, $\bm{\epsilon}$ is a normal Gaussian random variable, and $\odot$ denotes element-wise product operation.
In this way, the generated user latent variables $\bm{Z}_{u}^X$ only aggregate their interim representations $\widehat{\bm{U}}^X$, naturally aggregating the user homogeneous information from even-number-hop neighbors.

Building upon Eq.(\ref{vbge_1}), Eq.(\ref{vbge_2}) and Eq.(\ref{repara}), we obtain user latent variables $\bm{Z}_{u}^X$ in domain $X$. The item latent variables $\bm{Z}_{v}^X$ can also be obtained in similar learning processes. Meanwhile, the same learning process also holds for domain $Y$ with separate VBGEs. The overall generated user/item latent variables for domain $X/Y$ can be summarized as $\bm{Z}_{u}^X$, $\bm{Z}_{v}^X$, $\bm{Z}_{u}^Y$ and $\bm{Z}_{v}^Y$, respectively.

\subsection{Cross-Domain Information Regularizer} \label{Sec.Regularizer}
In this section, we introduce two kinds of regularizer to capture correlations across domains, aiming to learn unbiased representations with domain-shared information.
Specifically, the information bottleneck regularizer aims to capture the correlation between users and items across domains, while the contrastive information regularizer aims to capture overlapping user-user correlation across domains.

Before going on, we claim $\mathbf{X} ,\mathbf{X}^u, \mathbf{X}^v$ as the observed interaction information, user information, and item information in domain $X$, respectively.
Here the interaction information denotes all the observed user-item behaviors.
The user information term summarizes both the input users' features and their homogeneous structure (e.g., 2-hop user-user interaction), and the item information is defined similarly.
Based on two groups of users (e.g., domain $X$ contains overlapping user set $\mathcal{U}^o$ and non-overlapping user set $\mathcal{U}^x$), we divide the user representations $\bm{Z}_u^X \in \mathbb{R}^{|\mathcal{U}^X|\times F}$ into two groups: $\bm{Z}_u^{xo} \in \mathbb{R}^{|\mathcal{U}^o|\times F}$ and $\bm{Z}_u^x\in \mathbb{R}^{|\mathcal{U}^x|\times F}$ respectively.
For domain $Y$, we also claim similar symbols, and here we omit the detail for readability. 
Please refer to Table~\ref{tab::symbol} for the detail.

\begin{figure}[t]
	\begin{center}
		\includegraphics[width=8.5cm,height=4cm]{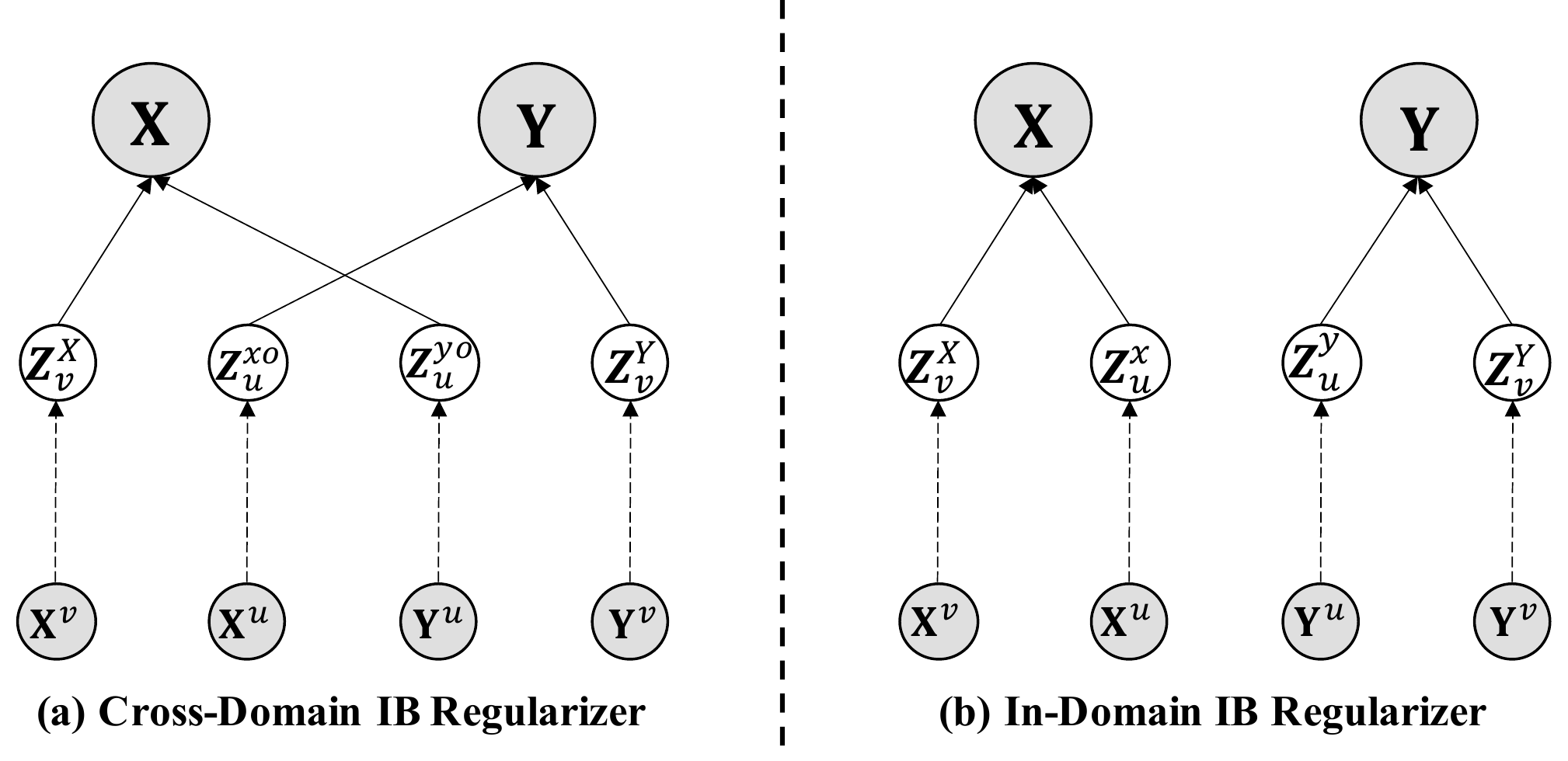}
		\caption{The graphical illustration of two IB regularizers. The $\bm{Z}_u^{xo},\bm{Z}_u^{yo}$ are overlapping user representations encoding the user information from different domains. The $\bm{Z}_u^{x},\bm{Z}_u^{y}$ are non-overlapped (cold-start) user representations, and $\bm{Z}_v^{X},\bm{Z}_v^{Y}$ are item representations. The $\mathbf{X}, \mathbf{Y}$ are detailed in Section \ref{Sec.Regularizer}.}
		\label{graphical}
	\end{center}
\end{figure}

\subsubsection{Information Bottleneck Regularizer}
To encourage the user/item representations to encode the domain-shared information, we adopt two information bottleneck regularizers: 
the cross-domain information bottleneck regularizer aims at modeling the cross-domain user-item interactions, while the in-domain information bottleneck regularizer aims at capturing the in-domain user-item interactions.
In this way, our method learns representations with domain-shared information rather than domain-specific pre-trained representations, and thereby has the capability to make recommendations in both domains.

\paragraph{\textbf{Cross-Domain Information Bottleneck Regularizer}}
In order to build cross-domain user-item interactions, the representations of the source domain should encode relevant information for direct prediction in the target domain.
For instance, given users in domain $X$ and items in domain $Y$, we hope that the user representations $\bm{Z}_u^{X}$ are capable to predict interactions among items $\bm{Z}_v^Y$ in domain $Y$.
Following this idea, we design the information bottleneck regularizer $\mathcal{L}_{o2Y}$ to impose the constraint (as shown in Fig.~\ref{graphical}(a)) on the overlapping users $\bm{Z}_u^{xo}$:
\begin{equation}
	\small
	\begin{split}
		\mathcal{L}_{o2Y} = & \beta_1 I(\bm{Z}_u^{xo};\mathbf{X}^u) - I(\bm{Z}_u^{xo};\mathbf{Y}) \\ 
		 & + \beta_2 I(\bm{Z}_v^Y;\mathbf{Y}^v) - I(\bm{Z}_v^{Y};\mathbf{Y}).\\
	\end{split}
	\label{overlapping}
\end{equation}
Here $\beta_1$, $\beta_2$ are tradeoff factors for domain $X$ and $Y$, and $\mathbf{X}^u$, $\mathbf{Y}^v$, $\mathbf{Y}$ denote the user information in domain $X$, item information in domain $Y$, and the interaction information in domain $Y$. For simplicity, we postulate the independence between $\bm{Z}_u^{xo}$ and $\bm{Z}_v^{Y}$, and then adopt the mutual information chain rule as:
\begin{equation}
	\small
	\begin{split}
		I(\bm{Z}_u^{xo};\mathbf{Y}) + I(\bm{Z}_v^{Y};\mathbf{Y}) = & I(\bm{Z}_u^{xo};\mathbf{Y}|\bm{Z}_v^{Y}) + I(\bm{Z}_v^{Y};\mathbf{Y}) \\
		= &I(\bm{Z}_u^{xo},\bm{Z}_v^{Y};\mathbf{Y}).\\
	\end{split}
	\label{chain}
\end{equation}
Finally, we can rewrite the Eq.(\ref{overlapping}) as follows:
\begin{equation}
	\small
	\begin{split}
		\mathcal{L}_{o2Y} =\underbrace{ \beta_1 I(\bm{Z}_u^{xo};\mathbf{X}^u) + \beta_2 I(\bm{Z}_v^Y;\mathbf{Y}^v)}_{\rm{Minimality}} - \underbrace{I(\bm{Z}_u^{xo},\bm{Z}_v^{Y};\mathbf{Y})}_{\rm{Reconstruction}}.\\
	\end{split}
	\label{overlapping2}
\end{equation}
Note that the minimality term penalizes the user/item information from their source domains, and the reconstruction term encourages representations to be predictive to reconstruct interactions in target domain $Y$. 
In other words, the regularizer encourages the user/item representations to encode the domain-shared information, and meanwhile limiting the domain-specific information.
This regularizer formulation also holds in the case of users from $Y$ and items from $X$, and finally leads to two regularizers $\mathcal{L}_{o2Y}$ and $\mathcal{L}_{o2X}$ for bi-directional CDR.
For the tractable objective function of Eq.(\ref{overlapping2}), please refer to Section~\ref{Sec.OptIB}.

\paragraph{\textbf{In-Domain Information Bottleneck Regularizer}}
To ensure the capability for making recommendations in each domain, we learn user/item representations with the unified information bottleneck manner.
Generally, we hope that the learned users $\bm{Z}_u^{X}$ and items $\bm{Z}_v^X$ ``forget'' certain redundant information (with $\mathbf{X}^u$ and $\mathbf{X}^v$), but remain the capability to reconstruct user-item interactions in their own domain.
Following this idea, we design the in-domain information bottleneck regularizer $\mathcal{L}_{x2X}$ to impose the constraint (as shown in Fig.~\ref{graphical}(b)) on non-overlapping users $\bm{Z}_u^{x}$:
\begin{equation}
	\small
	\begin{split}
		\mathcal{L}_{x2X} = \underbrace{\beta_1 I(\bm{Z}_u^{x};\mathbf{X}^u) + \beta_1 I(\bm{Z}_v^X;\mathbf{X}^v)}_{\rm{Minimality}} - \underbrace{I(\bm{Z}_u^{x},\bm{Z}_v^{X};\mathbf{X})}_{\rm{Reconstruction}}.\\
	\end{split}
	\label{nonoverlapping}
\end{equation}
Here $\beta_1$ is the tradeoff factor for domain $X$. The minimality term compresses the user/item representations with limited redundant information~\cite{vib}, and the reconstruction term encourages representations to remain predictive to reconstruct interactions in their own domain. This regularizer formulation also holds for domain $Y$, and finally leads to two regularizers $\mathcal{L}_{x2X}$ and $\mathcal{L}_{y2Y}$.
For the tractable objective function of Eq.(\ref{nonoverlapping}), please refer to Section~\ref{Sec.OptIB}.

\subsubsection{Contrastive Information Regularizer}
To further encourage the cross-domain user-user correlation, we devise the contrastive information regularizer.
Specifically, we refine the representations of overlapping users by measuring the mutual information between the representations $\bm{Z}_u^{xo}$ from domain $X$ and the $\bm{Z}_u^{yo}$ from domain $Y$. In this way, the user representations are enforced to capture the domain-shared information from both domains, thus deriving general representations for CDR. 
Thus, the regularizer can be formulated as:
\begin{equation}
	\small
	\begin{split}
		\mathcal{L}_{con} &= - \underbrace{I(\bm{Z}_u^{xo}; \bm{Z}_u^{yo})}_{\rm{Contrastive}} \\
		&= -I(\bm{Z}_u^{xo}; \bm{Z}_u^{yo}) + [H(\bm{Z}_u^{xo}|\mathbf{X}) - H(\bm{Z}_u^{xo}|\bm{Z}_u^{yo},\mathbf{X})] \\
		&= -I(\bm{Z}_u^{xo}; \bm{Z}_u^{yo}) + I(\bm{Z}_u^{xo}; \bm{Z}_u^{yo}|\mathbf{X}) \\
		&= -I(\bm{Z}_u^{xo}; \bm{Z}_u^{yo}; \mathbf{X}) \\
		&= -I(\bm{Z}_u^{xo}; \mathbf{X}) - I(\bm{Z}_u^{yo}; \mathbf{X}) + I(\bm{Z}_u^{xo}, \bm{Z}_u^{yo}; \mathbf{X}),
	\end{split}
	\label{contrastive}
\end{equation}
where $H(\cdot)$ means the information entropy.
For simplicity, we make the assumption that $q(\bm{Z}_u^{xo}|\mathbf{X})$ = $q(\bm{Z}_u^{xo}|\bm{Z}_u^{yo},\mathbf{X})$, and then adopt the mutual information chain rule.
As shown in Eq.(\ref{contrastive}), we can see two benefits in minimizing the $\mathcal{L}_{con}$:
(1) Minimizing the front two terms encourage that the latent variables $\bm{Z}_u^{xo}$ and $\bm{Z}_u^{yo}$ are informative to predict in domain $X$;
(2) Minimizing the last term penalizes the amount of \textit{jointly information}~\cite{iiae} between $\bm{Z}_u^{xo}, \bm{Z}_u^{yo}$ and the interaction information $\mathbf{X}$.
In other words, this regularizer enforces user representations from each domain to encode domain-shared information.
For tractable objective function of Eq.(\ref{contrastive}), please refer to Section~\ref{Sec.OptCIR}.

\subsection{Tractable Objective Function}

Actually, direct optimization of mutual information is intractable since it involves intractable integrals~\cite{vib,iiae,cvib}.
To optimize the proposed information regularizers, we derive their tractable objective functions with variational approximation~\cite{amortized} for computation. 
The overall objective function of CDRIB is presented at the end of this section.

\subsubsection{Optimizing the Information Bottleneck Regularizer} 
\label{Sec.OptIB}
To optimize information bottleneck regularizers in Eq.(\ref{overlapping2}) and Eq.(\ref{nonoverlapping}), we successively introduce the variational lower bound for the minimality terms and the reconstruction terms.

To optimize the \textbf{minimality term}, we measure it by the Kullback-Leibler (KL) divergence~\cite{klelbo} with variational approximation posterior distributions~\cite{vae}. Since all minimality terms have similar forms, we take term $I(\bm{Z}_u^{xo};\mathbf{X}^u)$ in Eq.(\ref{overlapping2}) as an example:
\begin{equation}
	\small
	\begin{split}
		I(\bm{Z}_u^{xo};\mathbf{X}^u) = & \mathbb{D}_{KL}\big(p_\theta(\bm{Z}_u^{xo}|\mathbf{X}^u)||p(\bm{Z}_u^{xo})\big). \\
	\end{split}
	\label{minimality}
\end{equation}
As most variational models adopted~\cite{vae,betavae,betatcvae}, we assume the prior distribution $p(\bm{Z}_u^{xo})$ as normal Gaussian distribution $\mathcal{N}\big(0, \text{diag}(\bm{I})\big)$.
To approximate true posterior distribution $p_\theta(\bm{Z}_u^{xo}|\mathbf{X}^u)$, similar with VGAE~\cite{vgae}, we also adopt neural graph encoder to generate the variational approximation posterior distribution $q_{\phi^X_u}(\bm{Z}_u^{xo}|\mathbf{X}^u)$.
Here we achieve the neural graph encoder by the proposed VBGE for users of domain $X$ with parameter $\phi^{X}_{u}$, since our VBGE generates $\bm{Z}_u^{xo}$ by aggregating homogeneous user information $\mathbf{X}^u$.
Hence, we minimize the minimality term by minimizing its approximation upper bound~\cite{elbomi} as the objective function:
\begin{equation}
	\small
	\begin{split}
		I(\bm{Z}_u^{xo};X^u)& \leq  \mathbb{D}_{KL}\big(q_{\phi^X_u}(\bm{Z}_u^{xo}|X^u)||p(\bm{Z}_u^{xo})\big) \\
		= & \mathbb{D}_{KL}\Big(\mathcal{N}\big(\bm{\mu}^{xo}_u, [\text{diag}(\bm{\sigma}^{xo}_u)]^2\big) || \mathcal{N}\big(0, \text{diag}(\bm{I})\big)\Big). \\
	\end{split}
	\label{minimality2}
\end{equation}
Note that the objective function also holds for minimality term $I(\bm{Z}_v^Y;\mathbf{Y}^v)$ in Eq.~(\ref{overlapping2}), and $I(\bm{Z}_u^{x};\mathbf{X}^u)$, $I(\bm{Z}_v^X;\mathbf{X}^v)$ in Eq.~(\ref{nonoverlapping}).

To optimize all \textbf{reconstruction terms}, as discussed by~\citeauthor{bvcf}~\cite{bvcf}, we also adopt variational approximation posterior distribution to render. The key idea is to reconstruct observed interactions with learned latent variables. For simplicity, we take $I(\bm{Z}_u^{xo},\bm{Z}_v^{Y};\mathbf{Y})$ in Eq.(\ref{overlapping2}) as an example:
\begin{equation}
	\small
	\begin{split}
		I(\bm{Z}_u^{xo},\bm{Z}_v^{Y};\mathbf{Y}) = \mathbb{E}_{p_\theta(\bm{Z}_u^{xo}|\mathbf{X}^u) p_\theta(\bm{Z}_v^{Y}|\mathbf{Y}^v)}[\log{p(\bm{A}^{Y}|\bm{Z}_u^{xo}, \bm{Z}_v^{Y})}].
	\end{split}
	\label{informative}
\end{equation}
where $\bm{A}^{Y}$ is the true user-item interaction matrix in domain $Y$.
Similarly, we adopt two graph neural networks, VBGE $q_{\phi^{X}_u}$ and $q_{\phi^{Y}_v}$, to depict their approximated distributions for $p_\theta(\bm{Z}_u^{xo}|\mathbf{X}^u)$ and $p_\theta(\bm{Z}_v^{Y}|\mathbf{Y}^v)$.
Then, we define a score function $s(\cdot)$ to measure the plausibility of user-item interactions.
Hence, we maximize the reconstruction term by maximizing the following lower bound:
\begin{equation}
	\small
	\begin{split}
		I(\bm{Z}_u^{xo}, \bm{Z}_v^{Y};\mathbf{Y})
		& \geq  \mathbb{E}_{q_{\phi^{X}_u}(\bm{Z}_u^{xo}|\mathbf{X}^u) q_{\phi^{Y}_v}(\bm{Z}_v^{Y}|\mathbf{Y}^v)}[\log{p(\bm{A}^{Y}|\bm{Z}_u^{xo}, \bm{Z}_v^{Y})}] \\
		= & \!\!\!\!\!\!\!\! \sum_{(u_i,v_j)\in \mathcal{E}^{Y}} \!\!\!\!\!\!\! \log\big(s(\bm{z}_{u_i}^{xo}, \bm{z}_{v_j}^y)\big) + \!\!\!\!\!\!\!\! \sum_{(u_i,\widetilde{v}_j)\notin \mathcal{E}^{Y}} \!\!\!\!\!\!\! \log\big(1 - s(\bm{z}_{u_i}^{xo}, \bm{z}_{\widetilde{v}_j}^y)\big),
	\end{split}
	\label{informative2}
\end{equation}
where $(u_i,v_j)$/$(u_i,\widetilde{v}_j)$ is the positive/negative user-item interaction pair. $\bm{z}_{u_i}^{xo}$, $\bm{z}_{v_j}^y$ and $\bm{z}^y_{\widetilde{v}_j}$ corresponds to the latent variables generated by VBGEs.
Here we implement the score function by the inner production followed by a sigmoid function, for its simplicity and effectiveness.
Note that the objective function also holds for reconstruction term $I(\bm{Z}_u^{x},\bm{Z}_v^{X};\mathbf{X})$ in Eq.~(\ref{nonoverlapping}).

\subsubsection{Optimizing the Contrastive Information Regularizer} 
\label{Sec.OptCIR}
To optimize the \textbf{contrastive term} of this regularizer, we follow the intuition from infomax~\cite{mine,dim} that utilizes neural networks to measure the contrastive mutual information~\cite{dgi,bigi,ya2021heterogeneous}.
Specifically, we define a discriminator $\mathcal{D}$ to measure the similarity of overlapping user latent variables from different domains, which discriminates $\bm{z}_{u_i}^{xo}$ from domain $X$ and $\bm{z}_{u_i}^{yo}$ from domain $Y$ with a similarity score. Hence, the lower bound of the contrastive term is as follows:
\begin{equation}
	\small
	\begin{split}
		I(\bm{Z}_u^{xo};\bm{Z}&_u^{yo}) =  \mathbb{E}_{p_\theta(\bm{Z}_{u}^{xo}|\mathbf{X}^u)p_\theta(\bm{Z}_{u}^{yo}|\mathbf{Y}^u)} [\log{\mathcal{D}(\bm{Z}_u^{xo}, \bm{Z}_u^{yo})}] \\
		\geq & \mathbb{E}_{q_{\phi^{X}_u}(\bm{Z}_{u}^{xo}|\mathbf{X}^u)q_{\phi^{Y}_u}(\bm{Z}_{u}^{yo}|\mathbf{Y}^u)} [\log{\mathcal{D}(\bm{Z}_u^{xo}, \bm{Z}_u^{yo})}] \\
		= & \!\!\!\!\!\!\!\!\!\!\!\!\!  \sum_{u_i, \widetilde{u}_i\in\mathcal{U}^o, \widetilde{u}_i \neq u_i} \!\!\!\!\!\!\!\!\!\!\!\!\!  \big[\!\log \big(\mathcal{D}(\bm{z}_{u_i}^{xo},\bm{z}_{u_i}^{yo})\big) \! + \! \log\big(1 - \mathcal{D}(\bm{z}_{u_i}^{xo},\bm{z}_{\widetilde{u}_i}^{yo})\big)\!\big].
	\end{split}
	\label{optcontrastive}
\end{equation}
where the $\bm{z}_{\widetilde{u}_i}^{yo}$ is a negative variable sampled from $\bm{Z}_{u}^{yo}$. Here we implement the discriminator $\mathcal{D}$ with a three-layer MLP followed by a sigmoid function, which can be denoted as:
\begin{equation}
	\small
	\begin{split}
		\mathcal{D}(\bm{z}_{u_i}^{xo},\bm{z}_{u_i}^{yo}) = \texttt{sigmoid}\big(\texttt{MLP}(\bm{z}_{u_i}^{xo}\oplus \bm{z}_{u_i}^{yo})\big),
	\end{split}
	\label{discriminator}
\end{equation}
where $\oplus$ denotes the vector concatenation operation. In this way, similar latent variables will enjoy higher similarity scores.

\begin{table*}[t]
\scriptsize
	\centering
	\caption{Statistics of four CDR scenarios. (\#Overlap denotes the number of overlapping users in the training set, \#Cold-start denotes the number of cold-start users in the validation and test set.)}
	\setlength\tabcolsep{15pt}{
		\begin{tabular}{l|cc|cc|ccc|c}
			\toprule
			Scenarios & $|\mathcal{U}|$ & $|\mathcal{V}|$ & Training  & \#Overlap &Validation & Test   & \#Cold-start & Density\\ 
			\midrule
			Music    & 50,841           & 43,858          & 674,233  & \multirow{2}{*}{15,081}  &19,837 & 19,670 &1,893 & 0.32\%\\
			Movie    & 87,875           & 38,643          & 1,127,424 & \multirow{2}{*}{} &28,589 & 28,876 & 1,885 & 0.34\%\\
			\midrule
			Phone    & 27,519           & 9,481           & 148,271 & \multirow{2}{*}{16,337} &6,417 & 6,322 & 2,049 & 0.61\%\\
			Elec     & 107,984                & 40,460          & 821,301  & \multirow{2}{*}{} &15,199 & 15,053 & 2,042 & 0.19\%\\
			\midrule
			Cloth    & 41,829            & 17,943          & 187,880  & \multirow{2}{*}{7,857} &3,156 & 3,085  & 990 & 0.25\%\\
			Sport    & 27,328          & 12,655          & 163,291   & \multirow{2}{*}{} &3,589 & 3,546  &981 & 0.49\%\\
			\midrule
			Game     & 25,025      & 12,319          & 155,036  & \multirow{2}{*}{1,737} &1,381 & 1,304  & 226 & 0.51\%\\
			Video    & 19,457        & 8,751           & 156,091   & \multirow{2}{*}{}  &1,435 & 1,458  & 217 & 0.93\%\\
			\bottomrule
		\end{tabular}
	}
	\label{dataset}
\end{table*}

\subsubsection{Optimizing the Overall Model} \label{Sec.OverallObj}
Based on the tractable objectives Eq.~(\ref{minimality2}), Eq.~(\ref{informative2}) and Eq.~(\ref{optcontrastive}), we can optimize the overall model in end-to-end framework.
In summary, we build the user-item interactions (e.g., Eq.~(\ref{overlapping2}) and Eq.~(\ref{nonoverlapping})) and the user-user correlations (e.g., Eq.~(\ref{contrastive})) across domains, and conclude final objective function as:
\begin{equation}
	\small
	\begin{split}
		\mathcal{L} = & \mathcal{L}_{x2X} + \mathcal{L}_{o2Y} + \mathcal{L}_{o2X} + \mathcal{L}_{y2Y} + \mathcal{L}_{con} \\
		= & \beta_1\big(I(\bm{Z}_u^X;\mathbf{X}^u) + I(\bm{Z}_v^X;\mathbf{X}^v)\big) \\ 
		& +\beta_2\big(I(\bm{Z}_u^Y;\mathbf{Y}^u) + I(\bm{Z}_v^Y;\mathbf{Y}^v)\big) \\
		& - I(\bm{Z}_u^{xo},\bm{Z}_v^{Y};\mathbf{Y}) - I(\bm{Z}_u^{x},\bm{Z}_v^{X};\mathbf{X}) \\
		& - I(\bm{Z}_u^{yo},\bm{Z}_v^{X};\mathbf{X}) - I(\bm{Z}_u^{y},\bm{Z}_v^{Y};\mathbf{Y}) \\
		& - I(\bm{Z}_u^{xo},\bm{Z}_u^{yo})\\
	\end{split}
	\label{loss}
\end{equation}
where the $\beta = \{\beta_1, \beta_2\}$ are hypeparameters used to penalize the learned representations biased in their training information. All minimality terms, reconstruction terms in the information bottleneck regularizers, and the contrastive term in the contrastive information regularizer can be tractably optimized by Eq.~(\ref{minimality2}), Eq.~(\ref{informative2}) and Eq.~(\ref{optcontrastive}), respectively.
Therefore, our method can be optimized with the mini-batch manner, and keeps efficiency time complexity mostly similar with other graph neural network methods~\cite{gcn,vgae,ngcf} as $\mathcal{O}((|\mathcal{E}^X| + |\mathcal{E}^Y|)F^2)$.

\section{Experiments}

\subsection{Datasets}
Following previous works, we adopt their selected cross-domain recommendation datasets~\cite{sscdr,tmcdr,conet,ppgn,bitg,ptucdr}, and the preprocessing settings~\cite{sscdr,tmcdr} to build our CDR scenarios.
Specifically, we conduct experiments on the large scale public Amazon~\cite{amazon} datasets\footnote{\url{http://jmcauley.ucsd.edu/data/amazon/index_2014.html}}.
The Amazon datasets consist of 24 disjoint item domains. Four pairs of domains: Music-Movie, Phone-Elec, Cloth-Sport, and Game-Video, are selected to evaluate CDR models for the bi-directional CDR scenarios.
Note that our goal is to recommend items in the target domain for those \textbf{(cold-start) users} only observed in the source domain.
Therefore, in the data preprocessing, we filter out the items that have fewer than 10 interactions and the users that have fewer than 5 interactions in their domains as previous works~\cite{tmcdr}, making the users/items access to learning representative embeddings from their source domain.
We also randomly select about 20\% overlapping users as cold-start users for test and validation (e.g., \textbf{the 10\% from Music to recommend in Movie and the residual 10\% from Movie to recommend in Music}) and the remaining users are used for training.
The concrete statistics of CDR scenarios are summarized in Table \ref{dataset}.

\subsection{Experiments setting}

\subsubsection{Evaluation Protocol}
Following previous works~\cite{sscdr,tmcdr}, we adopt the widely used \textit{leave-one-out evaluation} method to verify the effectiveness of methods.
For instance, given a ground truth interaction $(u_i,v_j)$ in domain $Y$, we first randomly select 999 items $\widetilde{v}_j$ from item set $\mathcal{V}^Y$ as negative samples.
Then, we calculate 1000 records (1 positive and 999 negative samples) by the learned representation $\bm{z}_{u_i}^X$ from domain $X$ and $\bm{z}_{v_j}^Y$/$\bm{z}_{\widetilde{v}_j}^Y$ from domain $Y$.
Next, we rank the record list and adopt three widely used recommendation metrics: MRR (mean reciprocal rank~\cite{mrr}), NDCG@\{5,10\}  (Normalized Discounted Cumulative Gain~\cite{ndcg}) and HR@\{1,5,10\} (Hit Rate) to show the \textit{top-k} recommendation performance.

\begin{table*}[t]
\scriptsize
\caption{Experimental results (\%) on the bi-directional Music-Movie CDR scenario. The best performance is \textbf{bold-faced} and the runner-up is \underline{underlined} in terms of the corresponding metric.}
\label{musicmovie}
\setlength\tabcolsep{0.8pt}{
{
\begin{tabular}{ccccccccccccc}
\toprule
\multirow{3}{*}{\bf Methods} & \multicolumn{6}{c}{\bf Music-domain recommendation} & \multicolumn{6}{c}{\bf Movie-domain recommendation}     \\
\cmidrule(r){2-7}\cmidrule{8-13}&
\multirow{2}{*}{MRR} &\multicolumn{2}{c}{NDCG} & \multicolumn{3}{c}{HR} & \multirow{2}{*}{MRR} &\multicolumn{2}{c}{NDCG} & \multicolumn{3}{c}{HR}\\
\cmidrule(r){3-4}\cmidrule(r){5-7}\cmidrule(r){9-10}\cmidrule{11-13} &  & @5 & @10  & @1  & @5  & @10  &  & @5  & @10  & @1  & @5  & @10   \\
\midrule

CML  & \phantom{0}4.19$\pm$0.09 &  \phantom{0}3.50$\pm$0.11    &  \phantom{0}4.53$\pm$0.14    &   \phantom{0}1.50$\pm$0.04 &  \phantom{0}5.50$\pm$0.17  &  \phantom{0}8.70$\pm$0.27 & 
\phantom{0}3.89$\pm$0.04   &  \phantom{0}2.91$\pm$0.04 & \phantom{0}3.95$\pm$0.03   &   \phantom{0}1.22$\pm$0.02  &   \phantom{0}4.62$\pm$0.05    &   \phantom{0}7.87$\pm$0.11  \\

BPRMF  & \phantom{0}4.51$\pm$0.03 &  \phantom{0}3.73$\pm$0.04    &  \phantom{0}4.72$\pm$0.04    &   \phantom{0}1.79$\pm$0.07 &  \phantom{0}5.63$\pm$0.13  &  \phantom{0}8.74$\pm$0.13 & 
\phantom{0}4.55$\pm$0.05   &  \phantom{0}3.58$\pm$0.07 & \phantom{0}4.72$\pm$0.08   &   \phantom{0}1.66$\pm$0.03  &   \phantom{0}5.55$\pm$0.12    &   \phantom{0}9.12$\pm$0.15  \\

NGCF  & \phantom{0}4.04$\pm$0.04 &  \phantom{0}3.18$\pm$0.06    &  \phantom{0}4.15$\pm$0.09    &   \phantom{0}1.33$\pm$0.04 &  \phantom{0}5.03$\pm$0.08  &  \phantom{0}8.86$\pm$0.10 & 
\phantom{0}4.08$\pm$0.08   &  \phantom{0}3.02$\pm$0.06 & \phantom{0}4.53$\pm$0.10   &   \phantom{0}0.92$\pm$0.06  &  \phantom{0}5.24$\pm$0.10    &   \phantom{0}9.92$\pm$0.16   \\

\midrule

CoNet  & \phantom{0}4.63$\pm$0.04 &  \phantom{0}3.94$\pm$0.05    &  \phantom{0}5.11$\pm$0.07    &   \phantom{0}1.90$\pm$0.01 &  \phantom{0}6.04$\pm$0.08  &  \phantom{0}9.27$\pm$0.15 & 
\phantom{0}4.17$\pm$0.08   &  \phantom{0}3.00$\pm$0.08 & \phantom{0}4.34$\pm$0.09   &   \phantom{0}1.07$\pm$0.06  &   \phantom{0}5.00$\pm$0.10    &   \phantom{0}9.17$\pm$0.15  \\

STAR  & \phantom{0}3.91$\pm$0.12 &  \phantom{0}3.12$\pm$0.12    &  \phantom{0}4.07$\pm$0.13    &   \phantom{0}1.43$\pm$0.09 &  \phantom{0}4.78$\pm$0.15  &  \phantom{0}7.71$\pm$0.20 & 
\phantom{0}3.91$\pm$0.18   &  \phantom{0}2.73$\pm$0.18 & \phantom{0}4.00$\pm$0.23   &   \phantom{0}0.89$\pm$0.12  &   \phantom{0}4.62$\pm$0.25    &   \phantom{0}8.60$\pm$0.04  \\

PPGN  & \phantom{0}4.18$\pm$0.04 &  \phantom{0}3.39$\pm$0.06    &  \phantom{0}4.55$\pm$0.07    &   \phantom{0}1.47$\pm$0.05 &  \phantom{0}5.32$\pm$0.11  &  \phantom{0}8.91$\pm$0.16 & 
\phantom{0}4.31$\pm$0.03   &  \phantom{0}3.19$\pm$0.02 & \phantom{0}4.65$\pm$0.04   &   \phantom{0}1.08$\pm$0.02  &   \phantom{0}5.44$\pm$0.07    &   10.00$\pm$0.10  \\

\midrule

EMCDR(CML)  & \phantom{0}3.80$\pm$0.01 &  \phantom{0}3.06$\pm$0.01    &  \phantom{0}3.97$\pm$0.02    &   \phantom{0}1.34$\pm$0.01 &  \phantom{0}4.78$\pm$0.01  &  \phantom{0}7.62$\pm$0.04 & 
\phantom{0}4.73$\pm$0.05   &  \phantom{0}3.68$\pm$0.05 & \phantom{0}4.97$\pm$0.06   &   \phantom{0}1.56$\pm$0.03  &   \phantom{0}5.81$\pm$0.08    &   \phantom{0}9.82$\pm$0.10  \\

EMCDR(BPRMF)  & \phantom{0}4.23$\pm$0.10 &  \phantom{0}3.46$\pm$0.11    &  \phantom{0}4.55$\pm$0.12    &   \phantom{0}1.55$\pm$0.03 &  \phantom{0}5.41$\pm$0.19  &  \phantom{0}8.79$\pm$0.18 & 
\phantom{0}4.49$\pm$0.07   &  \phantom{0}3.31$\pm$0.09 & \phantom{0}4.79$\pm$0.08   &   \phantom{0}1.18$\pm$0.04  &   \phantom{0}5.50$\pm$0.14    &   10.10$\pm$0.12  \\

EMCDR(NGCF)  & \phantom{0}4.36$\pm$0.08 &  \phantom{0}3.63$\pm$0.04    &  \phantom{0}4.70$\pm$0.08    &   \phantom{0}1.67$\pm$0.10 &  \phantom{0}5.61$\pm$0.09  &  \phantom{0}8.95$\pm$0.12 & 
\phantom{0}\underline{5.49$\pm$0.12}   &  \phantom{0}\underline{4.55$\pm$0.08} & \phantom{0}\underline{6.06$\pm$0.14}   &   \phantom{0}\underline{1.90$\pm$0.16}  &   \phantom{0}\underline{7.25$\pm$0.14}    &   \underline{11.89$\pm$0.10}  \\

SSCDR  & \phantom{0}1.95$\pm$0.01 &  \phantom{0}1.36$\pm$0.01    &  \phantom{0}1.82$\pm$0.01    &   \phantom{0}0.56$\pm$0.01 &  \phantom{0}2.15$\pm$0.02  &  \phantom{0}3.59$\pm$0.04 & 
\phantom{0}2.74$\pm$0.01   &  \phantom{0}1.99$\pm$0.01 & \phantom{0}2.72$\pm$0.01   &   \phantom{0}0.83$\pm$0.02  &   \phantom{0}3.19$\pm$0.03    &   \phantom{0}5.47$\pm$0.03  \\

TMCDR  & \phantom{0}\underline{4.78$\pm$0.04} &  \phantom{0}\underline{4.12$\pm$0.06}    &  \phantom{0}\underline{5.15$\pm$0.05}    &   \phantom{0}\underline{1.97$\pm$0.01} &  \phantom{0}\underline{6.27$\pm$0.12}  &  \phantom{0}\underline{9.48$\pm$0.08} & 
\phantom{0}5.29$\pm$0.05   &  \phantom{0}4.45$\pm$0.08 & \phantom{0}6.00$\pm$0.04   &   \phantom{0}1.71$\pm$0.08  &   \phantom{0}7.19$\pm$0.09    &   10.22$\pm$0.15  \\

SA-VAE  & \phantom{0}4.23$\pm$0.04 &  \phantom{0}3.44$\pm$0.04    &  \phantom{0}4.48$\pm$0.06    &   \phantom{0}1.53$\pm$0.02 &  \phantom{0}5.34$\pm$0.07  &  \phantom{0}8.57$\pm$0.15 & 
\phantom{0}5.25$\pm$0.09   &  \phantom{0}4.13$\pm$0.10 & \phantom{0}5.66$\pm$0.11   &   \phantom{0}1.58$\pm$0.05  &   \phantom{0}6.73$\pm$0.16    &   11.51$\pm$0.18  \\

\midrule

VBGE  & \phantom{0}4.30$\pm$0.06 &  \phantom{0}3.55$\pm$0.07    &  \phantom{0}4.60$\pm$0.09    &   \phantom{0}1.62$\pm$0.04 &  \phantom{0}5.53$\pm$0.16  &  \phantom{0}9.23$\pm$0.17 & 
\phantom{0}4.33$\pm$0.13   &  \phantom{0}3.25$\pm$0.12 & \phantom{0}4.80$\pm$0.17   &   \phantom{0}1.22$\pm$0.03  &   \phantom{0}5.35$\pm$0.24    &   10.07$\pm$0.39  \\

CDRIB  & \textbf{\phantom{0}7.01$\pm$0.05}* &  \textbf{\phantom{0}6.01$\pm$0.06}*    &  \textbf{\phantom{0}7.68$\pm$0.09}*    &   \textbf{\phantom{0}2.91$\pm$0.01}* &  \textbf{\phantom{0}9.08$\pm$0.15}*  &  \textbf{14.29$\pm$0.23}* & 
\textbf{\phantom{0}6.90$\pm$0.11}*   &  \textbf{\phantom{0}5.77$\pm$0.10}* & \textbf{\phantom{0}7.63$\pm$0.18}*   &   \textbf{\phantom{0}2.51$\pm$0.07}  &   \textbf{\phantom{0}9.07$\pm$0.14}*    &   \textbf{14.86$\pm$0.42}*  \\

\bottomrule

\end{tabular}
}}
\begin{center}
\normalsize* indicates that the improvements are statistically significant for $p < 0.05$ judged with the runner-up result in each case by paired t-test.
\end{center}
\end{table*}

\begin{table*}[t]
\scriptsize
\caption{Experimental results (\%) on the bi-directional Phone-Elec CDR scenario.}
\label{phoneelec}
\setlength\tabcolsep{0.8pt}{
{
\begin{tabular}{ccccccccccccc}
\toprule
\multirow{3}{*}{\bf Methods} & \multicolumn{6}{c}{\bf Phone-domain recommendation} & \multicolumn{6}{c}{\bf Elec-domain recommendation}     \\
\cmidrule(r){2-7}\cmidrule{8-13}&
\multirow{2}{*}{MRR} &\multicolumn{2}{c}{NDCG} & \multicolumn{3}{c}{HR} & \multirow{2}{*}{MRR} &\multicolumn{2}{c}{NDCG} & \multicolumn{3}{c}{HR}\\
\cmidrule(r){3-4}\cmidrule(r){5-7}\cmidrule(r){9-10}\cmidrule{11-13} &  & @5 & @10  & @1  & @5  & @10  &  & @5  & @10  & @1  & @5  & @10   \\
\midrule
CML  & \phantom{0}5.83$\pm$0.10 &  \phantom{0}5.14$\pm$0.11    &  \phantom{0}6.41$\pm$0.13    &   \phantom{0}2.55$\pm$0.05 &  \phantom{0}7.62$\pm$0.13  &  11.56$\pm$0.23 & 
\phantom{0}6.24$\pm$0.13   &  \phantom{0}5.54$\pm$0.17 & \phantom{0}6.85$\pm$0.12   &   \phantom{0}2.75$\pm$0.15  &   \phantom{0}8.34$\pm$0.23    &   12.44$\pm$0.10  \\

BPRMF  & \phantom{0}7.10$\pm$0.11 &  \phantom{0}\underline{6.71$\pm$0.11}    &  \phantom{0}7.65$\pm$0.07    &   \phantom{0}\underline{4.17$\pm$0.19} &  \phantom{0}9.05$\pm$0.08  &  11.98$\pm$0.22 & 
\phantom{0}6.79$\pm$0.20   &  \phantom{0}6.23$\pm$0.19 & \phantom{0}7.46$\pm$0.23   &   \phantom{0}3.42$\pm$0.18  &   \phantom{0}8.95$\pm$0.23    &   12.78$\pm$0.33  \\

NGCF  & \phantom{0}6.48$\pm$0.04 &  \phantom{0}5.82$\pm$0.08    &  \phantom{0}7.47$\pm$0.12    &   \phantom{0}2.48$\pm$0.10 &  \phantom{0}9.11$\pm$0.19  &  13.62$\pm$0.20 & 
\phantom{0}8.04$\pm$0.07   &  \phantom{0}7.41$\pm$0.08 & \phantom{0}9.09$\pm$0.11   &   \phantom{0}3.77$\pm$0.04  &   11.01$\pm$0.17    &   16.24$\pm$0.21  \\

\midrule

CoNet  & \phantom{0}\underline{7.18$\pm$0.18} &  \phantom{0}6.64$\pm$0.18    &  \phantom{0}7.68$\pm$0.20    &   \phantom{0}\textbf{4.45$\pm$0.15} &  \phantom{0}8.63$\pm$0.22  &  11.86$\pm$0.30 & 
\phantom{0}6.61$\pm$0.12   &  \phantom{0}5.97$\pm$0.11 & \phantom{0}7.19$\pm$0.14   &   \phantom{0}3.26$\pm$0.10  &   \phantom{0}8.52$\pm$0.13    &   12.33$\pm$0.25  \\

STAR  & \phantom{0}4.96$\pm$0.22 &  \phantom{0}4.22$\pm$0.20    &  \phantom{0}5.41$\pm$0.21    &   \phantom{0}1.87$\pm$0.19 &  \phantom{0}6.48$\pm$0.23  &  10.22$\pm$0.22 & 
\phantom{0}6.67$\pm$0.30   &  \phantom{0}5.86$\pm$0.32 & \phantom{0}7.42$\pm$0.34   &   \phantom{0}2.73$\pm$0.25  &   \phantom{0}8.92$\pm$0.41    &   13.76$\pm$0.47  \\

PPGN  & \phantom{0}6.51$\pm$0.09 &  \phantom{0}6.04$\pm$0.08    &  \phantom{0}7.39$\pm$0.16    &   \phantom{0}2.62$\pm$0.14 &  \phantom{0}9.49$\pm$0.16  &  13.60$\pm$0.23 & 
\phantom{0}8.07$\pm$0.03   &  \phantom{0}7.39$\pm$0.06 & \phantom{0}9.16$\pm$0.05   &   \phantom{0}3.78$\pm$0.03  &   10.87$\pm$0.10    &   16.38$\pm$0.13  \\

\midrule

EMCDR(CML)  & \phantom{0}5.69$\pm$0.01 &  \phantom{0}5.06$\pm$0.03    &  \phantom{0}6.32$\pm$0.02    &   \phantom{0}2.13$\pm$0.02 &  \phantom{0}7.96$\pm$0.04  &  11.89$\pm$0.05 & 
\phantom{0}7.28$\pm$0.01   &  \phantom{0}6.61$\pm$0.01 & \phantom{0}8.17$\pm$0.01   &   \phantom{0}3.29$\pm$0.01  &   \phantom{0}9.93$\pm$0.02    &   14.75$\pm$0.03  \\

EMCDR(BPRMF)  & \phantom{0}6.41$\pm$0.01 &  \phantom{0}5.74$\pm$0.01    &  \phantom{0}7.27$\pm$0.01    &   \phantom{0}2.58$\pm$0.02 &  \phantom{0}8.85$\pm$0.02  &  13.57$\pm$0.03 & 
\phantom{0}8.38$\pm$0.01   &  \phantom{0}7.70$\pm$0.01 & \phantom{0}9.45$\pm$0.02   &   \phantom{0}3.86$\pm$0.01  &   11.42$\pm$0.01    &   16.86$\pm$0.06  \\

EMCDR(NGCF)  & \phantom{0}6.93$\pm$0.04 &  \phantom{0}6.35$\pm$0.02    &  \phantom{0}\underline{7.73$\pm$0.03}    &   \phantom{0}3.11$\pm$0.06 &  \phantom{0}9.54$\pm$0.03  &  13.84$\pm$0.07 & 
\phantom{0}\underline{8.77$\pm$0.02}   &  \phantom{0}\underline{8.10$\pm$0.02} & \phantom{0}\underline{9.78$\pm$0.02}   &   \phantom{0}\underline{4.26$\pm$0.02}  &   \underline{11.76$\pm$0.03}    &   17.01$\pm$0.03  \\

SSCDR  & \phantom{0}3.33$\pm$0.01 &  \phantom{0}2.63$\pm$0.01    &  \phantom{0}3.33$\pm$0.01    &   \phantom{0}1.27$\pm$0.01 &  \phantom{0}3.97$\pm$0.02  &  6.14$\pm$0.03 & 
\phantom{0}4.94$\pm$0.01   &  \phantom{0}4.21$\pm$0.01 & \phantom{0}5.50$\pm$0.01   &   \phantom{0}2.19$\pm$0.02  &   \phantom{0}6.28$\pm$0.02    &   10.34$\pm$0.03  \\

TMCDR  & \phantom{0}6.74$\pm$0.06 &  \phantom{0}5.99$\pm$0.07    &  \phantom{0}7.56$\pm$0.09    &   \phantom{0}2.78$\pm$0.01 &  \phantom{0}9.10$\pm$0.18  &  13.95$\pm$0.22 & 
\phantom{0}8.23$\pm$0.03   &  \phantom{0}7.58$\pm$0.03 & \phantom{0}9.24$\pm$0.05   &   \phantom{0}3.78$\pm$0.03  &   11.23$\pm$0.04    &   16.38$\pm$0.08  \\

SA-VAE  & \phantom{0}6.85$\pm$0.15 &  \phantom{0}6.14$\pm$0.15    &  \phantom{0}7.66$\pm$0.18    &   \phantom{0}2.63$\pm$0.07 &  \phantom{0}\underline{9.65$\pm$0.19}  &  \underline{14.35$\pm$0.29} & 
\phantom{0}8.46$\pm$0.10   &  \phantom{0}7.74$\pm$0.12 & \phantom{0}9.49$\pm$0.13   &   \phantom{0}3.80$\pm$0.08  &   11.59$\pm$0.17    &   \underline{17.21$\pm$0.19}  \\

\midrule

VBGE  & \phantom{0}6.70$\pm$0.14 &  \phantom{0}6.02$\pm$0.16    &  \phantom{0}7.88$\pm$0.18    &   \phantom{0}2.54$\pm$0.20 &  \phantom{0}9.49$\pm$0.26  & 14.25$\pm$0.43  & 
\phantom{0}8.22$\pm$0.29   &  \phantom{0}7.91$\pm$0.40 & \phantom{0}9.44$\pm$0.34   &   \phantom{0}3.94$\pm$0.21  &   11.90$\pm$0.33    &   16.58$\pm$0.36  \\

CDRIB  & \textbf{\phantom{0}9.05$\pm$0.13}* &  \textbf{\phantom{0}8.19$\pm$0.15}*    &  \textbf{10.29$\pm$0.14}*    &   \phantom{0}4.00$\pm$0.12 &  \textbf{12.35$\pm$0.21}*  &  \textbf{18.87$\pm$0.30}* & 
\textbf{10.79$\pm$0.23}*   &\textbf{10.04$\pm$0.28}* & \textbf{12.15$\pm$0.29}*   &   \textbf{\phantom{0}5.32$\pm$0.13}*  &   \textbf{14.64$\pm$0.39}*    &   \textbf{21.21$\pm$0.44}*  \\
\bottomrule
\end{tabular}
}}
\end{table*}

\begin{table*}[!h]
\scriptsize
\caption{Experimental results (\%) on the bi-directional Cloth-Sport CDR scenario.}
\label{clothsport}
\setlength\tabcolsep{0.8pt}{
{
\begin{tabular}{ccccccccccccc}
\toprule
\multirow{3}{*}{\bf Methods} & \multicolumn{6}{c}{\bf Cloth-domain recommendation} & \multicolumn{6}{c}{\bf Sport-domain recommendation}     \\
\cmidrule(r){2-7}\cmidrule{8-13}&
\multirow{2}{*}{MRR} &\multicolumn{2}{c}{NDCG} & \multicolumn{3}{c}{HR} & \multirow{2}{*}{MRR} &\multicolumn{2}{c}{NDCG} & \multicolumn{3}{c}{HR}\\
\cmidrule(r){3-4}\cmidrule(r){5-7}\cmidrule(r){9-10}\cmidrule{11-13} &  & @5 & @10  & @1  & @5  & @10  &  & @5  & @10  & @1  & @5  & @10   \\
\midrule

CML  & \phantom{0}3.73$\pm$0.16 &  \phantom{0}3.19$\pm$0.19    &  \phantom{0}3.92$\pm$0.14    &\phantom{0}1.64$\pm$0.18  & \phantom{0}4.69$\pm$0.21 &  \phantom{0}6.97$\pm$0.11  &
\phantom{0}3.26$\pm$0.14   &  \phantom{0}2.65$\pm$0.15 & \phantom{0}3.29$\pm$0.16   &   \phantom{0}1.44$\pm$0.13  &   \phantom{0}3.82$\pm$0.17    &   \phantom{0}5.82$\pm$0.20  \\

BPRMF  & \phantom{0}3.02$\pm$0.18 &  \phantom{0}2.33$\pm$0.17    &  \phantom{0}3.26$\pm$0.15    &   \phantom{0}1.02$\pm$0.15 &  \phantom{0}3.80$\pm$0.20  &  \phantom{0}6.75$\pm$0.13 & 
\phantom{0}3.11$\pm$0.11   &  \phantom{0}2.50$\pm$0.15 & \phantom{0}3.16$\pm$0.15   &   \phantom{0}1.24$\pm$0.08  &   \phantom{0}3.72$\pm$0.27    &   \phantom{0}5.75$\pm$0.26  \\

NGCF  & \phantom{0}3.24$\pm$0.08 &  \phantom{0}2.50$\pm$0.08    &  \phantom{0}3.48$\pm$0.13    &   \phantom{0}1.01$\pm$0.06 &  \phantom{0}4.02$\pm$0.12  &  \phantom{0}7.07$\pm$0.30 & 
\phantom{0}3.42$\pm$0.06   &  \phantom{0}2.75$\pm$0.04 & \phantom{0}3.63$\pm$0.07   &   \phantom{0}1.17$\pm$0.06  &   \phantom{0}4.43$\pm$0.10    &   \phantom{0}7.22$\pm$0.11  \\

\midrule

CoNet  & \phantom{0}3.41$\pm$0.06 &  \phantom{0}2.66$\pm$0.11    &  \phantom{0}3.44$\pm$0.13    &   \phantom{0}1.32$\pm$0.05 &  \phantom{0}4.00$\pm$0.10  &  \phantom{0}6.42$\pm$0.22 & 
\phantom{0}3.44$\pm$0.09   &  \phantom{0}2.67$\pm$0.13 & \phantom{0}3.51$\pm$0.14   &   \phantom{0}1.39$\pm$0.11  &   \phantom{0}4.10$\pm$0.13    &   \phantom{0}6.46$\pm$0.17  \\

STAR  & \phantom{0}3.89$\pm$0.13 &  \phantom{0}3.33$\pm$0.11    &  \phantom{0}4.15$\pm$0.16    &   \phantom{0}1.39$\pm$0.10 &  \phantom{0}5.20$\pm$0.17  &  \phantom{0}7.76$\pm$0.27 & 
\phantom{0}3.00$\pm$0.16   &  \phantom{0}2.30$\pm$0.18 & \phantom{0}3.19$\pm$0.18   &   \phantom{0}0.80$\pm$0.16  &   \phantom{0}3.73$\pm$0.22    &   \phantom{0}6.49$\pm$0.22  \\

PPGN  & \phantom{0}3.34$\pm$0.09 &  \phantom{0}2.58$\pm$0.10    &  \phantom{0}3.57$\pm$0.04    &   \phantom{0}1.14$\pm$0.10 &  \phantom{0}4.05$\pm$0.16  &  \phantom{0}7.10$\pm$0.12 & 
\phantom{0}3.30$\pm$0.10   &  \phantom{0}2.54$\pm$0.08 & \phantom{0}3.58$\pm$0.11   &   \phantom{0}1.14$\pm$0.08  &   \phantom{0}4.52$\pm$0.10    &   \phantom{0}7.36$\pm$0.20  \\

\midrule

EMCDR(CML)  & \phantom{0}2.70$\pm$0.02 &  \phantom{0}1.96$\pm$0.01    &  \phantom{0}2.72$\pm$0.01    &   \phantom{0}0.91$\pm$0.01 &  \phantom{0}3.04$\pm$0.02  &  \phantom{0}5.42$\pm$0.06 & 
\phantom{0}2.91$\pm$0.02   &  \phantom{0}2.18$\pm$0.03 & \phantom{0}2.99$\pm$0.04   &   \phantom{0}0.92$\pm$0.02  &   \phantom{0}3.46$\pm$0.02    &   \phantom{0}6.02$\pm$0.06  \\

EMCDR(BPRMF)  & \phantom{0}4.30$\pm$0.01 &  \phantom{0}3.87$\pm$0.01    &  \phantom{0}4.48$\pm$0.03    &   \phantom{0}2.20$\pm$0.02 &  \phantom{0}5.41$\pm$0.03  &  \phantom{0}7.29$\pm$0.10 & 
\phantom{0}3.45$\pm$0.01   &  \phantom{0}2.69$\pm$0.02 & \phantom{0}3.71$\pm$0.04   &   \phantom{0}1.19$\pm$0.04  &   \phantom{0}4.24$\pm$0.07    &   \phantom{0}7.44$\pm$0.13  \\

EMCDR(NGCF)  & \phantom{0}\underline{4.89$\pm$0.06} &  \phantom{0}4.08$\pm$0.06    &  \phantom{0}\underline{5.17$\pm$0.08}    &   \phantom{0}\underline{2.68$\pm$0.04} &  \phantom{0}5.55$\pm$0.10  &  \phantom{0}7.91$\pm$0.15 & 
\phantom{0}\underline{4.07$\pm$0.09}   &  \phantom{0}\underline{3.24$\pm$0.10} & \phantom{0}\underline{4.03$\pm$0.12}   &   \phantom{0}1.46$\pm$0.09  &   \phantom{0}\underline{4.94$\pm$0.15}    &   \phantom{0}7.41$\pm$0.16  \\

SSCDR  & \phantom{0}2.96$\pm$0.01 &  \phantom{0}2.27$\pm$0.01    &  \phantom{0}3.06$\pm$0.04    &   \phantom{0}0.91$\pm$0.02 &  \phantom{0}3.64$\pm$0.03  &  \phantom{0}6.12$\pm$0.05 & 
\phantom{0}3.46$\pm$0.01   &  \phantom{0}2.66$\pm$0.01 & \phantom{0}3.75$\pm$0.02   &   \phantom{0}1.33$\pm$0.01  &   \phantom{0}4.18$\pm$0.02    &   \phantom{0}7.27$\pm$0.02  \\

TMCDR  & \phantom{0}4.84$\pm$0.10 &  \phantom{0}\underline{4.28$\pm$0.08}    &  \phantom{0}5.05$\pm$0.12    &   \phantom{0}2.62$\pm$0.07 &  \phantom{0}\underline{5.77$\pm$0.06}  &  \phantom{0}\underline{8.11$\pm$0.16} & 
\phantom{0}3.68$\pm$0.03   &  \phantom{0}3.14$\pm$0.03 & \phantom{0}3.84$\pm$0.04   &   \phantom{0}\underline{1.64$\pm$0.02}  &   \phantom{0}4.60$\pm$0.06    &   \phantom{0}7.18$\pm$0.07  \\

SA-VAE  & \phantom{0}4.43$\pm$0.11 &  \phantom{0}3.98$\pm$0.13    &  \phantom{0}4.59$\pm$0.08    &   \phantom{0}2.55$\pm$0.07 &  \phantom{0}5.34$\pm$0.17  &  \phantom{0}7.21$\pm$0.05 & 
\phantom{0}3.26$\pm$0.03   &  \phantom{0}2.71$\pm$0.03 & \phantom{0}3.72$\pm$0.02   &   \phantom{0}1.17$\pm$0.04  &   \phantom{0}4.29$\pm$0.04    &   \phantom{0}\underline{7.51$\pm$0.02}  \\

\midrule

VBGE  & \phantom{0}3.56$\pm$0.07 &  \phantom{0}2.76$\pm$0.05    &  \phantom{0}3.98$\pm$0.07    &   \phantom{0}1.27$\pm$0.08 &  \phantom{0}4.72$\pm$0.16  &  \phantom{0}7.54$\pm$0.11 & 
\phantom{0}3.68$\pm$0.10   &  \phantom{0}2.84$\pm$0.14 & \phantom{0}4.02$\pm$0.12   &   \phantom{0}1.21$\pm$0.09  &   \phantom{0}4.76$\pm$0.23    &   \phantom{0}7.87$\pm$0.22  \\

CDRIB  & \textbf{\phantom{0}6.29$\pm$0.29}* &  \textbf{\phantom{0}5.55$\pm$0.23}*    &  \textbf{\phantom{0}6.81$\pm$0.28}*    &   \textbf{\phantom{0}2.82$\pm$0.32} &  \textbf{\phantom{0}8.30$\pm$0.19}*  &  \textbf{12.19$\pm$0.29}* & 
\textbf{\phantom{0}5.59$\pm$0.12}*   &  \textbf{\phantom{0}4.89$\pm$0.10}* & \textbf{\phantom{0}6.22$\pm$0.09}*   &   \textbf{\phantom{0}2.08$\pm$0.09}  &   \textbf{\phantom{0}7.90$\pm$0.14}*    &   \textbf{12.04$\pm$0.20}*  \\
\bottomrule
\end{tabular}
}}
\end{table*}

\begin{table*}[t]
\scriptsize
\caption{Experimental results (\%) on the bi-directional Game-Video CDR scenario.}
\label{gamevideo}
\setlength\tabcolsep{0.8pt}{
{
\begin{tabular}{ccccccccccccc}
\toprule
\multirow{3}{*}{\bf Methods} & \multicolumn{6}{c}{\bf Game-domain recommendation} & \multicolumn{6}{c}{\bf Video-domain recommendation}     \\
\cmidrule(r){2-7}\cmidrule{8-13}&
\multirow{2}{*}{MRR} &\multicolumn{2}{c}{NDCG} & \multicolumn{3}{c}{HR} & \multirow{2}{*}{MRR} &\multicolumn{2}{c}{NDCG} & \multicolumn{3}{c}{HR}\\
\cmidrule(r){3-4}\cmidrule(r){5-7}\cmidrule(r){9-10}\cmidrule{11-13} &  & @5 & @10  & @1  & @5  & @10  &  & @5  & @10  & @1  & @5  & @10   \\
\midrule

CML  & \phantom{0}1.70$\pm$0.08 &  \phantom{0}1.10$\pm$0.08    &  \phantom{0}1.44$\pm$0.09    &   \phantom{0}0.43$\pm$0.08 &  \phantom{0}1.74$\pm$0.09  &  \phantom{0}2.82$\pm$0.18 & 
\phantom{0}1.46$\pm$0.14   &  \phantom{0}0.71$\pm$0.10 & \phantom{0}1.30$\pm$0.08   &   \phantom{0}0.13$\pm$0.17  &   \phantom{0}1.25$\pm$0.13    &   \phantom{0}3.07$\pm$0.10  \\

BPRMF  & \phantom{0}2.09$\pm$0.11 &  \phantom{0}1.25$\pm$0.15    &  \phantom{0}1.89$\pm$0.19    &   \phantom{0}0.65$\pm$0.07 &  \phantom{0}1.81$\pm$0.28  &  \phantom{0}3.77$\pm$0.40 & 
\phantom{0}2.44$\pm$0.26   &  \phantom{0}1.83$\pm$0.30 & \phantom{0}2.36$\pm$0.34   &   \phantom{0}0.83$\pm$0.28  &   \phantom{0}2.86$\pm$0.49    &   \phantom{0}4.46$\pm$0.56  \\

NGCF  & \phantom{0}\underline{2.94$\pm$0.07} &  \phantom{0}\underline{2.07$\pm$0.02}    &  \phantom{0}2.73$\pm$0.09    &   \phantom{0}1.07$\pm$0.03 &  \phantom{0}3.14$\pm$0.16  &  \phantom{0}5.14$\pm$0.22 & 
\phantom{0}3.72$\pm$0.05   &  \phantom{0}2.86$\pm$0.03 & \phantom{0}3.87$\pm$0.10   &   \phantom{0}1.57$\pm$0.06  &   \phantom{0}4.25$\pm$0.18    &   \phantom{0}7.41$\pm$0.18  \\

\midrule

CoNet  & \phantom{0}2.07$\pm$0.16 &  \phantom{0}1.54$\pm$0.15    &  \phantom{0}1.93$\pm$0.17    &   \phantom{0}0.79$\pm$0.16 &  \phantom{0}2.39$\pm$0.14  &  \phantom{0}3.69$\pm$0.22 & 
\phantom{0}2.47$\pm$0.18   &  \phantom{0}1.65$\pm$0.21 & \phantom{0}2.12$\pm$0.23   &   \phantom{0}0.69$\pm$0.10  &   \phantom{0}2.58$\pm$0.29    &   \phantom{0}4.04$\pm$0.38  \\

STAR  & \phantom{0}2.17$\pm$0.09 &  \phantom{0}1.61$\pm$0.04    &  \phantom{0}2.17$\pm$0.16    &   \phantom{0}0.57$\pm$0.03 &  \phantom{0}2.53$\pm$0.12  &  \phantom{0}4.27$\pm$0.40 & 
\phantom{0}2.62$\pm$0.10   &  \phantom{0}1.87$\pm$0.01 & \phantom{0}2.62$\pm$0.11   &   \phantom{0}0.72$\pm$0.08  &   \phantom{0}3.00$\pm$0.14    &   \phantom{0}5.30$\pm$0.18  \\

PPGN  & \phantom{0}2.77$\pm$0.06 &  \phantom{0}1.90$\pm$0.09    &  \phantom{0}\underline{2.84$\pm$0.10}    &   \phantom{0}0.92$\pm$0.08 &  \phantom{0}3.07$\pm$0.14  &  \phantom{0}\underline{5.98$\pm$0.32} & 
\phantom{0}3.90$\pm$0.11   &  \phantom{0}3.04$\pm$0.07 & \phantom{0}4.06$\pm$0.16   &   \phantom{0}1.44$\pm$0.07  &   \phantom{0}4.53$\pm$0.09    &   \phantom{0}7.68$\pm$0.32  \\

\midrule

EMCDR(CML)  & \phantom{0}2.12$\pm$0.01 &  \phantom{0}1.35$\pm$0.03    &  \phantom{0}1.82$\pm$0.04    &   \phantom{0}0.50$\pm$0.03 &  \phantom{0}2.17$\pm$0.07  &  \phantom{0}3.62$\pm$0.09 & 
\phantom{0}3.15$\pm$0.01   &  \phantom{0}2.57$\pm$0.01 & \phantom{0}3.23$\pm$0.02   &   \phantom{0}1.04$\pm$0.02  &   \phantom{0}4.25$\pm$0.04    &   \phantom{0}6.34$\pm$0.05  \\

EMCDR(BPRMF)  & \phantom{0}2.42$\pm$0.10 &  \phantom{0}1.56$\pm$0.14    &  \phantom{0}2.24$\pm$0.08    &   \phantom{0}0.65$\pm$0.03 &  \phantom{0}2.53$\pm$0.30  &  \phantom{0}4.63$\pm$0.13 & 
\phantom{0}\underline{4.15$\pm$0.06}   &  \phantom{0}\underline{3.29$\pm$0.04} & \phantom{0}4.29$\pm$0.05   &   \phantom{0}\underline{1.74$\pm$0.07}  &   \phantom{0}4.81$\pm$0.10    &   \phantom{0}7.94$\pm$0.32  \\

EMCDR(NGCF)  & \phantom{0}2.47$\pm$0.18 &  \phantom{0}1.86$\pm$0.15    &  \phantom{0}2.44$\pm$0.14    &   \phantom{0}0.50$\pm$0.05 &  \phantom{0}3.26$\pm$0.20  &  \phantom{0}5.07$\pm$0.17 & 
\phantom{0}3.98$\pm$0.01   &  \phantom{0}3.21$\pm$0.02 & \phantom{0}4.29$\pm$0.02   &   \phantom{0}1.53$\pm$0.06  &   \phantom{0}\underline{5.09$\pm$0.08}    &   \phantom{0}8.43$\pm$0.04  \\

SSCDR  & \phantom{0}1.98$\pm$0.03 &  \phantom{0}1.27$\pm$0.01    &  \phantom{0}1.59$\pm$0.03    &   \phantom{0}0.36$\pm$0.01 &  \phantom{0}1.96$\pm$0.03  &  \phantom{0}3.48$\pm$0.06 & 
\phantom{0}2.55$\pm$0.03   &  \phantom{0}1.86$\pm$0.02 & \phantom{0}2.61$\pm$0.02   &   \phantom{0}0.62$\pm$0.03  &   \phantom{0}3.14$\pm$0.06    &   \phantom{0}5.51$\pm$0.08  \\

TMCDR  & \phantom{0}2.52$\pm$0.05 &  \phantom{0}1.97$\pm$0.02    &  \phantom{0}2.58$\pm$0.07    &   \phantom{0}0.57$\pm$0.04 &  \phantom{0}3.40$\pm$0.05  &  \phantom{0}5.36$\pm$0.09 & 
\phantom{0}3.97$\pm$0.07   &  \phantom{0}3.15$\pm$0.07 & \phantom{0}\underline{4.41$\pm$0.08}   &   \phantom{0}1.46$\pm$0.07  &   \phantom{0}4.95$\pm$0.05    &   \phantom{0}\underline{8.85$\pm$0.11}  \\

SA-VAE  & \phantom{0}2.78$\pm$0.04 &  \phantom{0}1.83$\pm$0.02    &  \phantom{0}2.78$\pm$0.06    &   \phantom{0}\underline{1.11$\pm$0.03} &  \phantom{0}\underline{3.53$\pm$0.05}  &  \phantom{0}5.84$\pm$0.13 & 
\phantom{0}3.71$\pm$0.04   &  \phantom{0}2.66$\pm$0.05 & \phantom{0}3.71$\pm$0.06   &   \phantom{0}1.15$\pm$0.03  &   \phantom{0}4.20$\pm$0.11    &   \phantom{0}7.46$\pm$0.13  \\

\midrule

VBGE  & \phantom{0}2.99$\pm$0.10 &  \phantom{0}2.18$\pm$0.14    &  \phantom{0}2.78$\pm$0.16    &   \phantom{0}0.99$\pm$0.08 &  \phantom{0}3.45$\pm$0.31  &  \phantom{0}5.60$\pm$0.21 & 
\phantom{0}3.79$\pm$0.09   &  \phantom{0}2.95$\pm$0.12 & \phantom{0}3.93$\pm$0.14   &   \phantom{0}1.72$\pm$0.04  &   \phantom{0}4.39$\pm$0.17    &   \phantom{0}7.48$\pm$0.32  \\

CDRIB  & \textbf{\phantom{0}4.50$\pm$0.14}* &  \textbf{\phantom{0}3.61$\pm$0.13}*    &  \textbf{\phantom{0}4.58$\pm$0.13}*    &   \textbf{\phantom{0}1.68$\pm$0.11} &  \textbf{\phantom{0}5.44$\pm$0.25}*  &  \textbf{\phantom{0}8.51$\pm$0.30}* & 
\textbf{\phantom{0}5.65$\pm$0.19}*   &  \textbf{\phantom{0}4.65$\pm$0.19}* & \textbf{\phantom{0}6.49$\pm$0.14}*   &   \textbf{\phantom{0}1.85$\pm$0.34}  &   \textbf{\phantom{0}7.41$\pm$0.31}*    &   \textbf{13.17$\pm$0.29}*  \\

\bottomrule
\end{tabular}
}}
\end{table*}

\subsubsection{Compared Methods}
To verify the effectiveness of cross-domain recommendation to cold-start users, we compare CDRIB with the following state-of-the-art baselines which can be divided into two branches.

\textit{Single-domain recommendation}:
In this branch, we merge all interactions of both domains as a single domain, and then apply the following widely-used CF-based methods:
(1) \textbf{CML}~\cite{cml} learns the user and item representations under the metric learning idea, which exploits L2 distance and supposes that a user representation has a smaller distance to the interacted items than the items not interacted yet.
(2) \textbf{BPRMF}~\cite{bprmf} is a well-known method in recommender systems, which further adopts the inner dot semantic metric to measure the relevance between users and items.
(3) \textbf{NGCF}~\cite{ngcf} is a typical graph neural network based approach to learn user and item representations which stacks three-layer GCNs~\cite{gcn} to explore high-order connectivity information between users and items.
(4) \textbf{VBGE} is a degenerate version of CDRIB, which replaces all regularizers with VGAE~\cite{vgae} loss function to evaluate our variational bipartite graph encoder for recommendation.
	
\textit{Cross-domain recommendation}:
Cross-domain baselines learn different user/item representations for each domain separately.
We first adopt several typical cross-domain models without considering the user cold-start problem as baselines:
(1) \textbf{CoNet}~\cite{conet} first models interactions of two domains with two separate feed-forward networks, and then transfer knowledge with a cross-connection network between the two base networks.
(2) \textbf{STAR}~\cite{star} devises an encoder topology structure, and introduces a shared base network for all domains to transfer knowledge.
(3) \textbf{PPGN}~\cite{ppgn} learns user/item representations by two separate GCNs with a shared user embedding layer.
Then, we employ current state-of-the-art EMCDR-based methods that aim at the user cold-start recommendation, as strong baselines:
(4) \textbf{EMCDR}~\cite{emcdr} is the first work that presents a pipeline paradigm for CDR to cold-start users.
Considering the impact of the pre-trained user/item representations to the EMCDR performance, we adopt three different pre-training approaches, including `CML', `BPRMF' and `NGCF', to initialize the user/item representations in EMCDR, termed as `\textbf{EMCDR(CML)}', `\textbf{EMCDR(BPRMF)}' and `\textbf{EMCDR(NGCF)}', respectively.
(5) \textbf{SSCDR}~\cite{sscdr} is an expanded EMCDR-based method, which first pre-trains representations via the `CML' approach, and then learns the cross-domain mapping function via multi-hop neighboring information of overlapping users.
(6) \textbf{TMCDR}~\cite{tmcdr} is also an expanded EMCDR-based method, which utilizes the learned representations of `BPRMF' and trains a meta cross-domain mapping function following the meta-learning framework~\cite{meta}.
(7) \textbf{SA-VAE}~\cite{recsys} is a variational EMCDR-based method, which first utilizes VAE framework~\cite{vae} to produce user latent variables for each domain, and then trains the mapping function for prediction.
Since all of these methods (1-6) conduct CDR from source to target domain in one direction, we run two times to achieve bi-directional CDR.
Note that we do not compare our method with other methods~\cite{catn,dcdir} that exploit additional side information for fair comparisons.


\subsubsection{Parameter Settings}
In our experiments, except the SSCDR implemented by ourselves, we use official implementations of other baselines.
For all the methods, we adopt the same value for the common hyperparameters, including: the embedding dimension $F$ to 128, the batch size to 1024.
For the specific hyperparameters in the baselines, we use the values reported in their original literature.
Besides, for all the EMCDR variants, we employ the same \texttt{MLP} architecture as suggested in EMCDR (i.e., embedding dimension $[F \rightarrow 2\times F \rightarrow F]$).
For our model CDRIB, we tune the Lagrangian multiplier $\beta_1$ and $\beta_2$ in \{0.5, 1.0, 1.5, 2.0\}, the dropout rate in \{0.1, 0.2, 0.3, 0.4\}, the L2 regularizer weight in \{0.001, 0.0005, 0.0001\}, the slope of LeakyReLU is fixed as 0.1, the learning rate in \{0.01, 0.005, 0.001\}, and the number of graph encoder layer in \{1,2,3,4\}. 
As suggested by NGCF~\cite{ngcf} and LightGCN~\cite{lightgcn}, we concatenate the graph encoder output of each layer as the final output of the graph encoder.
We use Adam~\cite{adam} optimizer for our model while for baselines, we keep the original optimizers reported in their literature.
For all these methods, we run each experiment with a random seed for five times and select the best result according to the highest MRR performance on the validation set, tuned by grid search.

\subsection{Performance Comparisons}
Table \ref{musicmovie}-\ref{gamevideo} report the mean and standard deviation results of MRR, NDCG and HR on the four CDR scenarios.
From them we have several observations:
\subsubsection{Comparison with Single-Domain Models}
For single-domain methods, we have the following observations: 
(1) Graph-neural-network-based model NGCF and VBGE consistently outperform CML and BPRMF in the term of HR@10.
This observation validates that capturing the multi-hop neighboring information is to some extent helpful for learning user and item representations.
(2) Our VBGE outperforms NGCF in many cases, which indicates that capturing the homogeneous information is beneficial to learn representations in the bipartite graph.
(3) The EMCDR-based cross-domain methods show satisfied improvement than their single-domain counterparts. 
This is mainly because those single-domain methods fail to consider the difference between domains, making it hard to recognize the transferring information for the cold-start users.
This observation demonstrates that modeling the difference between domains and capturing the valuable transferring information is necessary for CDR.

\subsubsection{Comparison with Cross-Domain Models}
For cross-domain methods, we have the following observations:
(1) CoNet, STAR and PPGN show comparable results with single-domain methods. The reason might be that those methods focus on transferring information to enhance the overlapping user recommendations, and easily generate biased representations for cold-start users.
(2) For EMCDR-based methods, EMCDR(CML), EMCDR(BPRMF) and EMCDR(NGCF) always show better performance than its single-domain counterparts CML, BPRMF and NGCF, respectively.
It verifies that the EMCDR-based paradigm is a promising way to make recommendations for cold-start users.
(3) EMCDR(NGCF) reaches superior performance to EMCDR(BPRMF) and EMCDR(CML), which demonstrates that the initialized representations could significantly affect the recommendation quality under the EMCDR-based paradigm.
(4) According to the Phone-domain results (in Table \ref{phoneelec}) and the Game-domain results (in Table \ref{gamevideo}), the improvement of EMCDR-based methods are very limited and even worse than its single-domain counterparts in some cases, which indicates that a simple function may be not enough to describe the complex mapping relation of cross-domain user representations.
(5) To promote the mapping function capability, TMCDR additionally introduces the meta-learning techniques and SA-VAE applies the variational inference framework to further attain better recommendation performance than EMCDR.
However, SSCDR also aims to enhance the mapping function of EMCDR(CML) by considering user/item multi-hop neighboring information.
In our experiments, SSCDR shows worse results than EMCDR(CML) except the Cloth-Sport CDR scenario, we assume the reason to the fact that SSCDR can be sensitive to the initialized pre-trained representations.
(6) Comparing with all EMCDR-based baselines, our CDRIB achieves large improvements on all CDR scenarios with all metrics, which demonstrates that learning a mapping function on the biased representations can be hard to achieve optimal results.
In contrast, our CDRIB captures the difference of domains and utilizes the IB and contrastive regularizers to encourage representations to focus on domain-shared information and compress the domain-specific information. 
In this way, the generated cold-start user representations are unbiased on each single domain and can predict in the target domain directly.

\subsection{Ablation Study}
To investigate the effectiveness of our model components, we conduct two variants of CDRIB (the results are shown in Table~\ref{ablation}).
Specifically, the \textit{w/o} Con variant is a degenerate version of CDRIB without the contrastive regularizer, and the \textit{w/o} In-IB\&Con variant is that without both the in-domain bottleneck regularizer and the contrastive regularizer.
Note that we remain the cross-domain IB regularizer in each variant to maintain the capability of making cross-domain recommendations.
From Table \ref{ablation}, we can observe that:
(1) Comparing the \textit{w/o} In-IB\&Con with cross-domain baselines, \textit{w/o} In-IB\&Con can gain superior performance over them, which indicates that our joint learning paradigm to capture the domain-shared information is more powerful than the pipeline paradigm of EMCDR.
(2) The \textit{w/o} Con variant consistently outperforms than \textit{w/o} In-IB\&Con, which demonstrates the in-domain IB regularizer can further encourage those non-overlapping (cold-start) user representations to encode the domain-shared information, which could enhance our model generalization for CDR to cold-start users.
(3) Additionally, our CDRIB yields satisfactory improvements than \textit{w/o} Con, where we suppose the reason that the contrastive regularizer can explicitly align the overlapping user representations, which could enhance the user-level cross-domain correlations.

\begin{table}[t]
\scriptsize
	\centering
	\caption{Performance comparison (\%) of different variants. \#\textit{w/o} Con denotes the variant without the contrastive regularizer, \#\textit{w/o} In-IB\&Con denotes the variant without both in-domain information bottleneck and contrastive regularizers.}
	\setlength\tabcolsep{7pt}{
		\begin{tabular}{lcccc}
			\toprule
			\multirow{2}{*}{Scenarios} & \multirow{2}{*}{Metrics} & \multicolumn{3}{c}{Model Variants}                   \\ \cline{3-5} & & \#\textit{w/o} In-IB\&Con & \#\textit{w/o} Con & CDRIB \\
			\midrule
			
			\multirow{3}{*}{Music}     & MRR                          & \phantom{0}5.89                              & \phantom{0}6.59 & \phantom{0}7.01 \\
			\multirow{3}{*}{}          & NDCG@10                        & \phantom{0}6.58                              & \phantom{0}7.21 & \phantom{0}7.68 \\
			\multirow{3}{*}{}          & HR@10                        & 12.37                              & 13.68 & 14.29 \\
			\cline{2-5}
			\specialrule{0em}{2pt}{2pt}
			\multirow{3}{*}{Movie}     & MRR                          & \phantom{0}5.56                              & \phantom{0}6.42 & \phantom{0}6.90 \\
			\multirow{3}{*}{}          & NDCG@10                        & \phantom{0}6.17                              & \phantom{0}7.04 & \phantom{0}7.63 \\
			\multirow{3}{*}{}          & HR@10                        & 12.67                              & 13.88 & 14.86 \\
			\midrule
			
			\multirow{3}{*}{Phone}     & MRR                          & \phantom{0}6.20                              & \phantom{0}7.98 & \phantom{0}9.05 \\
			\multirow{3}{*}{}          & NDCG@10                        & \phantom{0}6.98                              & \phantom{0}9.00 & 10.29 \\
			\multirow{3}{*}{}          & HR@10                        & 13.48                              & 16.87 & 18.87 \\
			\cline{2-5}
			\specialrule{0em}{2pt}{2pt}
			\multirow{3}{*}{Elec}     & MRR                          & \phantom{0}9.44                              & 10.32 & 10.79 \\
			\multirow{3}{*}{}          & NDCG@10                        & 10.49                              & 11.51 & 12.15 \\
			\multirow{3}{*}{}          & HR@10                        & 18.49                              & 19.92 & 21.21 \\
			\midrule

			\multirow{3}{*}{Cloth}     & MRR                          & \phantom{0}4.28                              & \phantom{0}5.26 & \phantom{0}6.29 \\
			\multirow{3}{*}{}          & NDCG@10                        & \phantom{0}4.69                              & \phantom{0}5.77 & \phantom{0}6.81 \\
			\multirow{3}{*}{}          & HR@10                        & \phantom{0}9.89                              & 11.15 & 12.19 \\
			\cline{2-5}
			\specialrule{0em}{2pt}{2pt}
			\multirow{3}{*}{Sport}     & MRR                          & \phantom{0}3.88                              & \phantom{0}4.78 & \phantom{0}5.59 \\
			\multirow{3}{*}{}          & NDCG@10                        & \phantom{0}4.20                              & \phantom{0}5.32 & \phantom{0}6.22 \\
			\multirow{3}{*}{}          & HR@10                        & \phantom{0}9.09                              & 10.69 & 12.04 \\
			\midrule

			\multirow{3}{*}{Game}     & MRR                          & \phantom{0}3.63                              & \phantom{0}4.18 & \phantom{0}4.50 \\
			\multirow{3}{*}{}          & NDCG@10                        & \phantom{0}3.78                              & \phantom{0}4.12 & \phantom{0}4.58 \\
			\multirow{3}{*}{}          & HR@10                        & \phantom{0}7.27                              & \phantom{0}8.05 & \phantom{0}8.51 \\
			\cline{2-5}
			\specialrule{0em}{2pt}{2pt}
			\multirow{3}{*}{Video}     & MRR                          & \phantom{0}4.15                              & \phantom{0}4.78 & \phantom{0}5.65 \\
			\multirow{3}{*}{}          & NDCG@10                        & \phantom{0}4.36                              & \phantom{0}5.18 & \phantom{0}6.49 \\
			\multirow{3}{*}{}          & HR@10                        & \phantom{0}9.33                              & 10.77 & 13.17 \\

			\bottomrule
		\end{tabular}
	}
	\label{ablation}
\end{table}

\subsection{Discussions of the Overlapping User Proportion}
To investigate the robustness of our model regarding the number of overlapping users that bridge different domains, we conduct experiments with different available proportions of the overlapping users.
Table~\ref{trainingrate} reports the CDRIB and SA-VAE recommendation performances of the target domain in the corresponding cross-domain scenario (e.g., the target Music domain with source domain Movie) in terms of 20\%, 40\%, 60\%, 80\% and 100\% ratio of overlapping users for training.
From Table~\ref{trainingrate}, we have the following observations: 
(1) With the increase of the overlapping user training ratio, the recommendation performance steadily improves, which demonstrates that training overlapping users is effective to enhance the correlation across domains.
(2) Our model shows robust performance to make recommendations for cold-start users than the strongest baseline, SA-VAE. 
Furthermore, even in the 20\% ratio setting, our CDRIB is still better than other baselines shown in Table~\ref{musicmovie}-\ref{gamevideo}.
The phenomenon reveals that our CDRIB is capable of effectively capturing the domain-shared information even in the case that there are only a few overlapping users in the training procedure.

\subsection{Discussions of the Cold-Start User Interaction Number}
To investigate the effectiveness of our model for the target cold-start (non-overlapping) users regarding the number of observed interactions in their \textbf{source domain}, we further conduct experiments on groups with different amounts of source interactions.
As shown in Table~\ref{interactionnumber}, we select five user groups in terms of the interaction number in the source domain, and report the target domain results for each CDR scenario (e.g., we report the average performance in the target Movie domain of those users with 5-10 interactions available in source Music domain).
From Table~\ref{interactionnumber}, we have the following observations: 
(1) In many cases, with the increasing number of observed interactions, the performances of CDRIB and SA-VAE are gradually improving. 
(2) Interestingly, for all domains, we observe the performance fluctuations with the number of the source interactions increase in some cases.
We assume the reason might be that more interactions in the source domain may not always contribute more valuable clues to capture the domain-shared information. 
Taking the romantic movie recommendation as an example, observed romantic music interactions may provide valuable clues to recommend romantic movies, but observed rock music may not be helpful for the romantic movie recommendation. 
(3) Our model shows more effective recommendation performance in all user groups than SA-VAE. Besides, even in the worst user group for each CDR scenario, our CDRIB are shown better performance than other baselines in Table~\ref{musicmovie}-\ref{gamevideo}. This observation also demonstrates that our CDRIB is effective to make cold-start recommendations even in the case that the users contain few observed interactions in the source domain.

 \begin{table}[t]
 \scriptsize
	\centering
	\caption{The robustness of models regarding the overlapping user proportion. \#Ratio denotes the proportion of the training overlapping users in observed overlapping user set.}
	\setlength\tabcolsep{1pt}{{
		\begin{tabular}{ccccccccc}
			\toprule
			\multirow{2}{*}{Scenarios} & \multirow{2}{*}{\#Ratio} & \multicolumn{3}{c}{CDRIB} & \multicolumn{3}{c}{SA-VAE}                   \\ \cmidrule(r){3-5} \cmidrule(r){6-8} & &MRR & NDCG@10 & HR@10  &MRR & NDCG@10 & HR@10 \\
			\midrule
			
			\multirow{5}{*}{Movie $\to$ Music}     & 20 \%                          & \phantom{0}6.13                              & \phantom{0}6.52 & 12.71 & \phantom{0}2.98                   & \phantom{0}2.94 & \phantom{0}5.60 \\
			\multirow{5}{*}{}          & 40 \%                        & \phantom{0}6.34                              & \phantom{0}6.99 & 13.07  &\phantom{0}3.36              & \phantom{0}3.42 & \phantom{0}6.33\\
			\multirow{5}{*}{}          & 60 \%                       & \phantom{0}6.55                              & \phantom{0}7.17 & 13.42 &\phantom{0}3.96                              & \phantom{0}3.89 & \phantom{0}7.53\\
			\multirow{5}{*}{}          & 80 \%                       & \phantom{0}6.87                              & \phantom{0}7.45 & 14.01 & \phantom{0}4.13                              & \phantom{0}4.09 & \phantom{0}7.82\\
			\multirow{5}{*}{}          & 100 \%                       & \phantom{0}7.01                              & \phantom{0}7.68 & 14.29 & \phantom{0}4.23                              & \phantom{0}4.48 & \phantom{0}8.57\\
			\midrule

			\multirow{5}{*}{Elec $\to$ Phone}     & 20 \%                          & \phantom{0}7.99                              & \phantom{0}9.09 & 17.02 & \phantom{0}6.24    & \phantom{0}7.03 & 13.04\\
			\multirow{5}{*}{}          & 40 \%                        & \phantom{0}8.08                              & \phantom{0}9.20 & 17.10 & \phantom{0}6.51    & \phantom{0}7.23 & 13.57\\
			\multirow{5}{*}{}          & 60 \%                       & \phantom{0}8.40                              & \phantom{0}9.57 & 17.76 & \phantom{0}6.63    & \phantom{0}7.47 & 13.82\\
			\multirow{5}{*}{}          & 80 \%                       & \phantom{0}8.82                              & \phantom{0}9.89 & 18.55 &\phantom{0}6.77    & \phantom{0}7.52 & 14.01\\
			\multirow{5}{*}{}          & 100 \%                       & \phantom{0}9.05                             & 10.29 & 18.87 & \phantom{0}6.85  & \phantom{0}7.66 & 14.35\\
			\midrule

			\multirow{5}{*}{Sport $\to$ Cloth}     & 20 \%                          & \phantom{0}3.65                              & \phantom{0}3.98 & \phantom{0}7.71 & \phantom{0}2.49                              & \phantom{0}2.75 & \phantom{0}6.82\\
			\multirow{5}{*}{}          & 40 \%                        & \phantom{0}4.45                              & \phantom{0}4.12 & \phantom{0}8.21 & \phantom{0}3.13                              & \phantom{0}3.55 & \phantom{0}7.28\\
			\multirow{5}{*}{}          & 60 \%                       & \phantom{0}5.26                              & \phantom{0}4.80 & \phantom{0}9.53 & \phantom{0}3.60                              & \phantom{0}3.67 & \phantom{0}7.69\\
			\multirow{5}{*}{}          & 80 \%                       & \phantom{0}6.05                             & \phantom{0}6.25 & 11.26 & \phantom{0}3.92                              & \phantom{0}4.64 & \phantom{0}7.77\\
			\multirow{5}{*}{}          & 100 \%                       & \phantom{0}6.29                              & \phantom{0}6.81 & 12.19 & \phantom{0}4.43    & \phantom{0}4.59 & \phantom{0}7.21\\
			\midrule
			
			\multirow{5}{*}{Video $\to$ Game}     & 20 \%                          & \phantom{0}3.33                              & \phantom{0}3.29 & \phantom{0}6.21 & \phantom{0}2.29    & \phantom{0}2.25 & \phantom{0}4.31\\
			\multirow{5}{*}{}          & 40 \%                        & \phantom{0}3.45                              & \phantom{0}3.43 & \phantom{0}7.13 & \phantom{0}2.44    & \phantom{0}2.35 & \phantom{0}4.57\\
			\multirow{5}{*}{}          & 60 \%                       & \phantom{0}3.94                              & \phantom{0}3.84 & \phantom{0}7.90 & \phantom{0}2.55    & \phantom{0}2.80 & \phantom{0}5.34\\
			\multirow{5}{*}{}          & 80 \%                       & \phantom{0}4.22                              & \phantom{0}4.24 & \phantom{0}8.14 & \phantom{0}2.80    & \phantom{0}2.82 & \phantom{0}5.83\\
			\multirow{5}{*}{}          & 100 \%                       & \phantom{0}4.50                              & \phantom{0}4.58 & \phantom{0}8.51 & \phantom{0}2.78  & \phantom{0}2.78 & \phantom{0}5.84\\
			
			\bottomrule
		\end{tabular}
	}}
	\label{trainingrate}
\end{table}

\subsection{Hyperparameter Analysis}
\begin{figure}[t]
	\begin{center}
		\includegraphics[width=8.7cm,height=7cm]{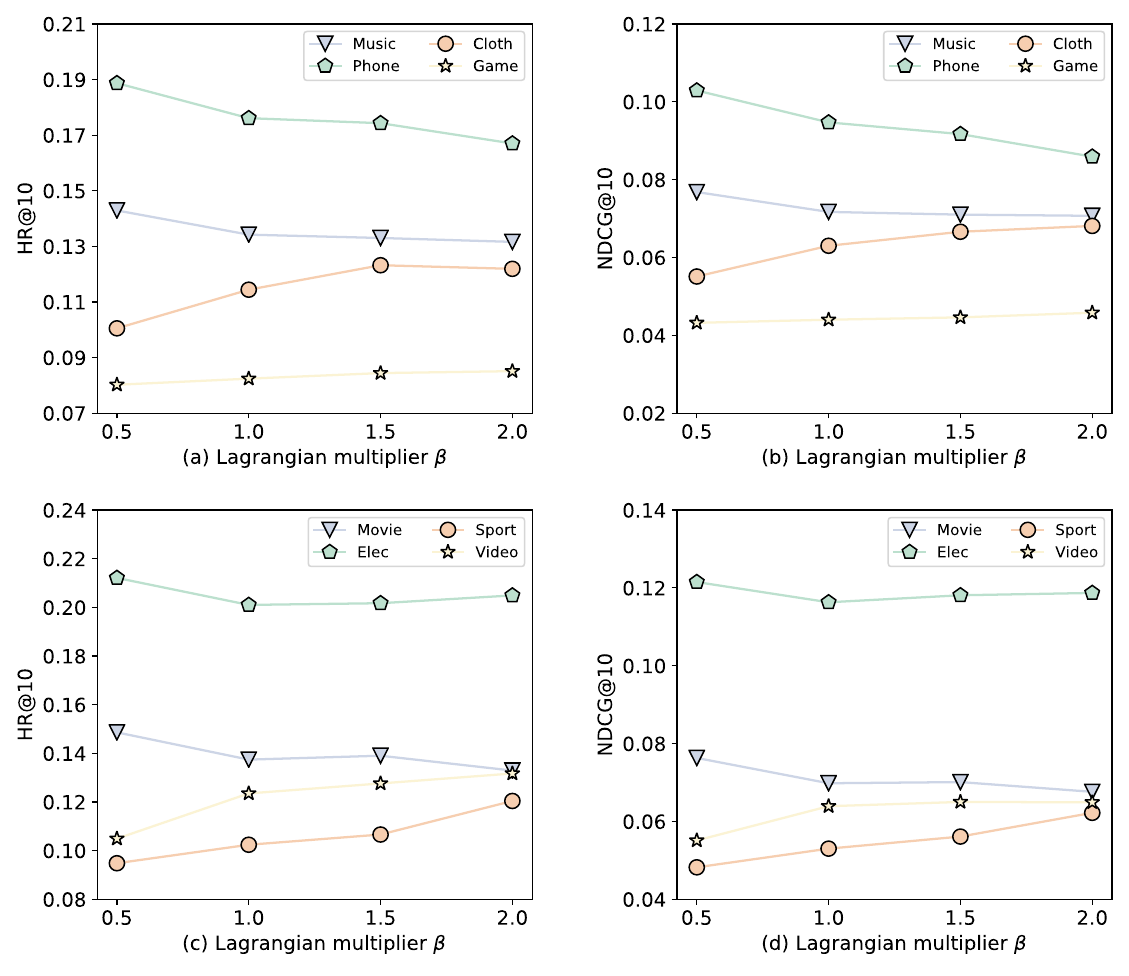}
		\caption{Effect of the Lagrangian multiplier.}
		\label{parameter_beta}
	\end{center}
\end{figure}

\begin{figure}[t]
	\begin{center}
		\includegraphics[width=8.7cm,height=7cm]{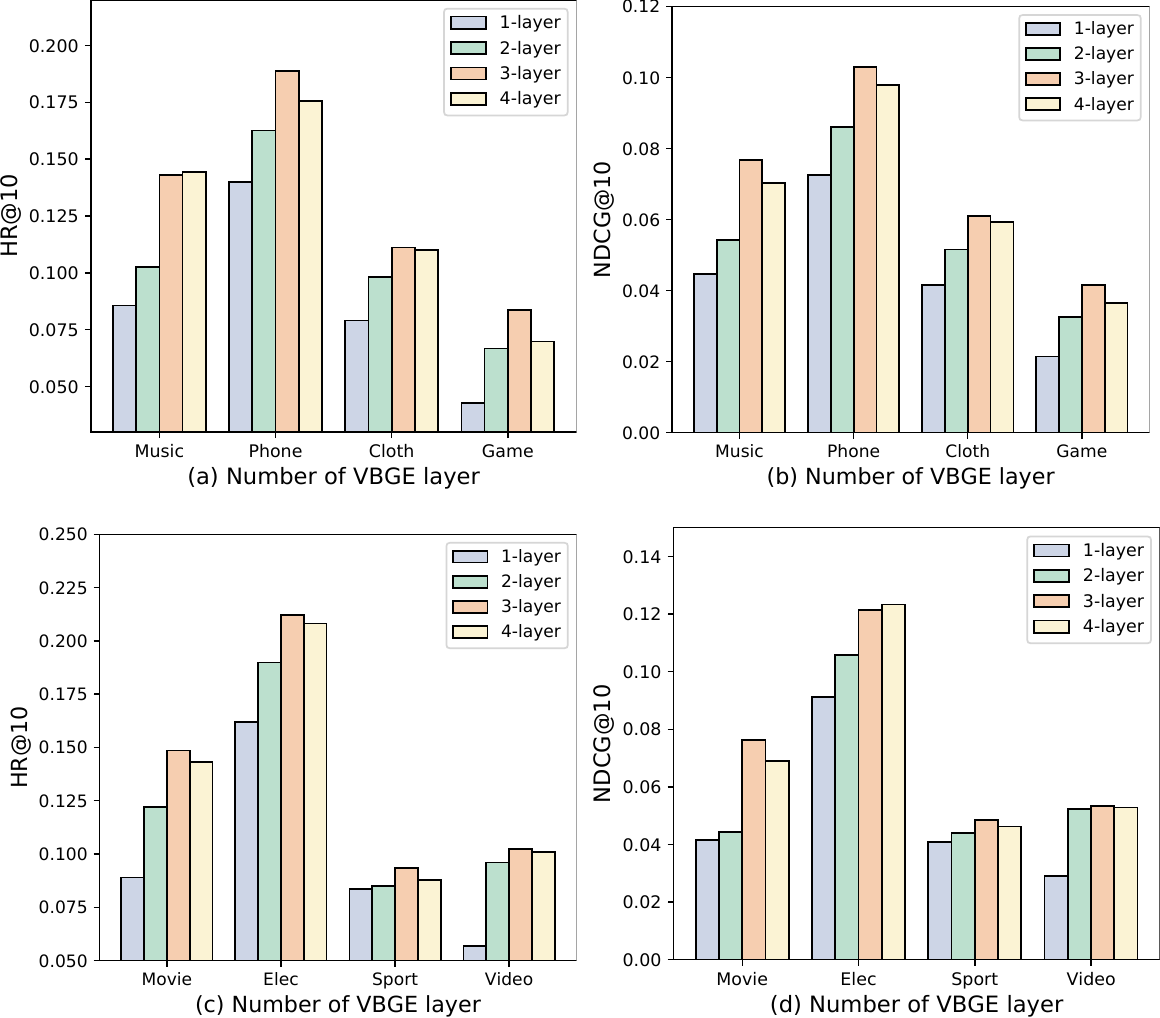}
		\caption{Impact of VBGE layer number.}
		\label{parameter_gnn}
	\end{center}
\end{figure}

We analyze the effect of two important hyperparameters: the Lagrangian multiplier and the layer number of VBGE:
(1) Fig.~\ref{parameter_beta} shows the effect of the Lagrangian multiplier.
For simplicity, we set the same weight $\beta$ for both the $\beta_1$ and $\beta_2$ in Eq.(\ref{loss}), and select from 0.5 to 2.0 with the step length 0.5.
According to Fig.~\ref{parameter_beta} with Table~\ref{dataset}, we observe that the best $\beta$ is associated with the scenario interaction scale, and $\beta$ tends to have a relatively smaller value for the scenarios with more interactions, such as the Music-Movie scenario and the Phone-Elec scenario.
The reason might be that the more interaction scenario contains more complex interaction patterns, and thus requires a smaller Lagrangian multiplier value to preserve the capability of user/item representations.
(2) Fig.~\ref{parameter_gnn} shows the impact of the layer number of VBGE.
Based on them, we find that aggregating the neighborhood information is significantly helpful to enhance the recommendation results of the cold-start users.
Nevertheless, the 4-layer variant results are lower than the 3-layer variant in many cases. We attribute the reason to the over-smoothing issue~\cite{graphsurvey} in graph neural networks~\cite{cui2020edge}.

\begin{table}[t]
 \scriptsize
	\centering
	\caption{The effectiveness of models regarding the number of observed interactions in source domain. \#Inter denotes the training number of user interactions from their source domain.}
	\setlength\tabcolsep{1pt}{{
		\begin{tabular}{cccccccc}
			\toprule
\multirow{2}{*}{Scenarios} & \multirow{2}{*}{\#Inter} & \multicolumn{3}{c}{CDRIB} & \multicolumn{3}{c}{SA-VAE}  \\ \cmidrule(r){3-5} \cmidrule(r){6-8} & &MRR & NDCG@10 & HR@10  &MRR & NDCG@10 & HR@10 \\

			\midrule
			
			\multirow{5}{*}{Music $\to$ Movie }     & 5-10                           & \phantom{0}6.18                              & \phantom{0}6.68 & 12.83 &\phantom{0}5.10   & \phantom{0}5.22 & 10.92\\ 
			\multirow{5}{*}{}          & 11-20                         & \phantom{0}6.68                              & \phantom{0}7.20 & 13.83 &\phantom{0}5.37   & \phantom{0}5.64 & 10.77\\
			\multirow{5}{*}{}          & 21-30                        & \phantom{0}7.70                              & \phantom{0}8.48 & 16.48 &\phantom{0}6.88   & \phantom{0}6.31 & 13.89\\
			\multirow{5}{*}{}          & 31-40                        & \phantom{0}7.57                              & \phantom{0}8.57 & 16.54 &\phantom{0}5.08   & \phantom{0}5.96 & 11.03\\
			\multirow{5}{*}{}          & 41-50                        & \phantom{0}6.83                              & \phantom{0}7.74 & 16.42 &\phantom{0}6.26   & \phantom{0}6.11 & 13.79\\
			\midrule

			\multirow{5}{*}{Phone $\to$ Elec}     & 5-10                           & 10.95                              & 12.44 & 21.75 & \phantom{0}8.74   & \phantom{0}9.94 & 17.81\\
			\multirow{5}{*}{}          & 11-20                         & 10.52                              & 11.78 & 20.85 &\phantom{0}8.44   & \phantom{0}9.58 & 17.02\\
			\multirow{5}{*}{}          & 21-30                        & \phantom{0}9.16                              & \phantom{0}9.83 & 17.06 &\phantom{0}6.56   & \phantom{0}6.99 & 12.08\\
			\multirow{5}{*}{}          & 31-40                        & \phantom{0}9.43                              & 10.03 & 17.96 &\phantom{0}5.50   & \phantom{0}5.46 & \phantom{0}9.15\\
			\multirow{5}{*}{}          & 41-50                        & 15.97                              & 16.81 & 25.33 &\phantom{0}8.21   & \phantom{0}8.24 & 11.42\\
			\midrule

			\multirow{5}{*}{Cloth $\to$ Sport}     & 5-10                           & \phantom{0}5.26                              & \phantom{0}5.80 & 11.05 & \phantom{0}3.58   & \phantom{0}3.87 & \phantom{0}7.75\\
			\multirow{5}{*}{}          & 11-20                         & \phantom{0}6.05                              & \phantom{0}7.06 & 14.13 & \phantom{0}2.92   & \phantom{0}3.05 & \phantom{0}6.33\\
			\multirow{5}{*}{}          & 21-30                        & \phantom{0}6.99                              & \phantom{0}7.03 & 14.13 & \phantom{0}3.21   & \phantom{0}3.70 & \phantom{0}7.54\\
			\multirow{5}{*}{}          & 31-40                        & \phantom{0}6.65                              & \phantom{0}7.01 & 11.11 & \phantom{0}3.04   & \phantom{0}3.85 & \phantom{0}6.88\\
			\multirow{5}{*}{}          & 41-50                        & \phantom{0}7.33                              & \phantom{0}10.67 & 13.09 & \phantom{0}3.15   & \phantom{0}3.33 & \phantom{0}7.14\\
			\midrule

			\multirow{5}{*}{Game $\to$ Video}     & 5-10                           & \phantom{0}4.95                              & \phantom{0}5.29 & 11.32  & \phantom{0}3.61                              & \phantom{0}3.74 & \phantom{0}7.47\\
			\multirow{5}{*}{}          & 11-20                         & \phantom{0}8.28                              & \phantom{0}9.01 & 16.98 & \phantom{0}4.82                              & \phantom{0}4.76 & \phantom{0}8.78\\
			\multirow{5}{*}{}          & 21-30                        & 11.38                              & 12.40 & 21.10 & \phantom{0}3.24                              & \phantom{0}2.93 & \phantom{0}4.72\\
			\multirow{5}{*}{}          & 31-40                        & \phantom{0}7.45                              & 10.40 & 23.91 & \phantom{0}3.57                              & \phantom{0}4.01 & \phantom{0}9.28\\
			\multirow{5}{*}{}          & 41-50                        & 13.06                              & 17.43 & 36.36 & \phantom{0}4.69   & \phantom{0}4.46 & \phantom{0}6.59\\

			\bottomrule
		\end{tabular}
	}}
	\label{interactionnumber}
\end{table}

\section{Related Work}
\label{related}

\subsection{Cross-Domain Recommendation}
Generally, cross-domain~\cite{debiasing,icde} recommendation leverages observed interactions from a relative source domain to improve the recommendation performance in the target domain.
According to the aiming group of users, recent efforts of CDR can be roughly divided into the following two categories.
\subsubsection{Cross-Domain Recommendation to Overlapping Users}
To alleviate the data sparsity issue, recent studies~\cite{cmf,cdcf,conet} extend CF-based methods to learn better user representations with interactions from different domains.
Concretely, CMF~\cite{cmf} and CDCF~\cite{cdcf} leverage the source domain interactions as additional features for overlapping users to achieve performance improvement in the target domain.
DARec~\cite{darec} transfers knowledge across domains by relying on adversarial domain adaptive technique.
CoNet~\cite{conet} and DDTCDR~\cite{ddtcdr} are two feed-forward neural network methods, which first use two base networks to model different domains and then transfer the learned overlapping user representations across two base networks.
PPGN~\cite{ppgn} and Bi-TGCF~\cite{bitg} are two graph neural network approaches, which stack several GCN layers to jointly model the interactions from two domains to learn more robust representations.
Besides, the multi-domain methods such as STAR~\cite{star} and MMoE~\cite{mmoe} could also be adapted in this branch, they utilize shared encoders to transfer information for in-domain recommendation.
Aside from this group of researches, our paper leverages the overlapping users to bridge different domains, and makes recommendations for cold-start users in the target domains, having completely different purposes from this group of researchers.

\subsubsection{Cross-Domain Recommendation to Cold-Start Users}
To mitigate the cold-start issue, several works~\cite{emcdr,sscdr,dcdir} explore correlations between the source domain and the target domain in different aspects.
For example, CBMF~\cite{cbmf} devises a cluster-level cross-domain matrix to learn the correlation between user clusters and item clusters, and CATN~\cite{catn} proposes an end-to-end framework to capture the aspect-level correlation between user reviews and item descriptions across domains.
Recently, EMCDR~\cite{emcdr} develops an effective Embedding-and-Mapping paradigm to capture the correlation between user representations, which first pre-trains user/item representations in the source domain and then directly learns a mapping function to transfer user representations into the target domain for cold-start users.
Following this paradigm, SSCDR~\cite{sscdr} exploits multi-hop information of users to improve the robustness of the mapping function, and DCDIR~\cite{dcdir} further introduces meta-path~\cite{homohete} information of item knowledge graph~\cite{wang2017knowledge,sheng2020adaptive} to fuse the external knowledge.
Recent TMCDR~\cite{tmcdr} and PTUPCDR~\cite{ptucdr} follow the MAML~\cite{maml} framework to learn a meta network to substitute the mapping function for the better recommendation, which can still be regarded as a particular form of the embedding-and-mapping paradigm.
Besides, motivated by the success of VAE~\cite{vae} framework in collaborative filtering, CDVAE~\cite{vcross}, AlignVAE~\cite{linkedvae} and SA-VAE~\cite{recsys} are proposed based on Bayesian-VAE to learn the mapping function.
Compared with these aforementioned methods, our CDRIB has important design differences as follows:
%
(1) We devise a variational graph encoder, capturing the global structural information of the user-item interaction graph.
(2) These works mostly employ biased user/item representations of each domain, while we leverage the IB principle to learn domain-shared representations. 


\subsection{Information Bottleneck}
Information bottleneck is a basic concept of information theory~\cite{orignalib}, which has been employed in many fields such as computer vision~\cite{betatcvae} and neural language processing~\cite{vibbert}.
Specifically, VIB~\cite{vib} first adopts IB to deep neural networks with variational inference framework~\cite{amortized}, and shows more robustness and expressiveness of learned representations.
Besides, $\beta$-VAE~\cite{betavae} further demonstrates that IB can also be used to learn disentanglement representations~\cite{disenib, lin2021disentangled}.
Recently, several works~\cite{gib, cvib} have proposed to extend IB for graph data, such as GIB~\cite{gib} which utilizes IB to regularize the structural information in the node classification task, CVIB~\cite{cvib} which devises a counterfactual~\cite{counterfactual} objective with IB for the recommendation,
and $L_0$-SIGN~\cite{sign} exploits IB to identify useful interaction for graph classification task.
Although those methods successfully apply IB for many tasks, but they are not designed for CDR, which is challenging and remains an unexplored problem.

\section{Conclusion}

In this paper, we propose a novel approach CDRIB for CDR to cold-start users via variational information bottleneck.
To the best of our knowledge, this paper is the first work to capture the domain-shared information for cold-start users.
CDRIB first devises the variational bipartite graph encoder to generate user/item representations, and then proposes two information bottleneck regularizers to build user-item correlations across domains.
With an additional contrastive information regularizer to model user-user correlations across domains, our CDRIB could capture the domain-shared information for user/items and limit domain-specific information.
Extensive experiments demonstrate that CDRIB outperforms current state-of-the-art methods on four CDR scenarios, indicating that capturing the domain-shared information is beneficial for CDR to cold-start users.
Besides, detailed analyses demonstrate the effectiveness of model components and the robustness of CDRIB in various CDR scenarios.
In the future, we would further investigate information bottleneck for the multi-domain recommendation and other CDR scenarios.

\section*{Acknowledgement}
We would like to thank anonymous reviewers for their constructive comments.
This work was supported by the National Key Research and Development Program of China under Grant No.2021YFB3100600, the Strategic Priority Research Program of Chinese Academy of Sciences under Grant No.XDC02040400, and the Youth Innovation Promotion Association of CAS under Grant No.2021153.

\bibliographystyle{IEEEtranN}
\bibliography{IEEEexample.bib}

\end{document}